\documentclass[10pt,letterpaper]{article}
\usepackage[top=0.85in,footskip=0.75in]{geometry}

\usepackage{pdfpages}
\usepackage{graphicx}
\date{}

\newcommand{\wue}{$^{3,4}$}
\newcommand{\ks }{$^{12,4}$}
\newcommand{\unit}[1]{\ensuremath{\, \mathrm{#1}}}

\begin{document}
\vspace*{0.35in}

\begin{flushleft}
{\Large
\textbf\newline{Participatory Patterns in an International Air Quality Monitoring Initiative}
}
\newline
\\
Alina S\^irbu$^{1,2,\ast}$, Martin Becker\wue{}, Saverio Caminiti$^5$, Bernard De Baets$^{13}$, Bart Elen$^{6}$, Louise Francis$^7$, Pietro Gravino$^{8}$, Andreas Hotho\wue{}, Stefano Ingarra$^{1}$, Vittorio Loreto$^{9,1,10}$, Andrea Molino$^{11}$, Juergen Mueller\ks{}, Jan Peters$^{6}$, Ferdinando Ricchiuti$^{11}$, Fabio Saracino$^{1}$, Vito D.~P.~Servedio$^{5,9}$, Gerd Stumme\ks{}, Jan Theunis$^{6}$, Francesca Tria$^1$, Joris Van den Bossche$^{6,13}$
\\
\bigskip
\bf{1} Complex Networks and Systems Lagrange Laboratory, Institute for Scientific Interchange Foundation, Turin, Italy
\\
\bf{2} Department of Computer Science and Engineering, University of Bologna, Italy
\\
\bf{3} Department for Artificial Intelligence and Applied Computer Science, University of W\"urzburg, W\"urzburg, Germany
\\
\bf{4} L3S Research Center, Gottfried Wilhelm Leibniz Universit\"at Hannover, Hannover, Germany
\\
\bf{5} Institute for Complex Systems (ISC), CNR, Rome, Italy
\\
\bf{6} VITO - Flemish Institute for Technological Research, Mol, Belgium
\\
\bf{7} Extreme Citizen Science Research Group, Department of Civil, Environmental and Geomatic Engineering, University College London, London, United Kingdom
\\
\bf{8} Physics Department, University of Bologna, Bologna, Italy
\\
\bf{9} Physics Department, Sapienza University, Rome, Italy
\\
\bf{10} SONY-CSL Computer Science Lab, Paris, France
\\
\bf{11} CSP - Innovation in ICT, Torino, Italy
\\
\bf{12} Department of Electrical Engineering and Computer Science, University of Kassel, Kassel, Germany
\\
\bf{13} KERMIT, Dept. of Mathematical Modelling, Statistics and Bioinformatics, Faculty of Bioscience Engineering, Ghent University, Ghent, Belgium
\\
\bigskip

* alina.sirbu@unibo.it

\end{flushleft}
\section*{Abstract}
The issue of sustainability is at the top of the political and societal agenda, being considered of extreme importance and urgency. Human individual action impacts the environment both locally (e.g., local air/water quality, noise disturbance) and globally (e.g., climate change, resource use). Urban environments represent a crucial example, with an increasing realization that the most effective way of producing a change is involving the citizens themselves in monitoring campaigns (a citizen science bottom-up approach). This is possible by developing novel technologies and IT infrastructures enabling large citizen participation. Here, in the wider framework of one of the first such projects, we show results from an international competition where citizens were involved in mobile air pollution monitoring using low cost sensing devices, combined with a web-based game to monitor perceived levels of pollution. Measures of shift in perceptions over the course of the campaign are provided, together with insights into participatory patterns emerging from this study. Interesting effects related to inertia and to direct involvement in measurement activities rather than indirect information exposure are also highlighted, indicating that direct involvement can enhance learning and environmental awareness. In the future, this could result in better adoption of policies towards decreasing pollution.


\section*{Introduction}


Air pollution has an important effect on our health, with an increasing number of studies showing higher risk of respiratory and cardiovascular diseases for people exposed to higher pollution levels\cite{Lancet2013,Lave2013}.
 In this context, keeping air pollution at bay has been a major priority for policy makers in the past decades. 
A lot of effort has been put into monitoring and controlling air pollution. Large scale monitoring networks routinely monitor target pollutants. They allow for temporal trends in air pollution to be tracked. Significant effort has also been made to make information accessible to the wider public. However, several papers indicate that official monitoring networks do not have sufficient spatial coverage to provide detailed information on personal exposure of people, as for some pollutants, this may vary substantially among micro-environments\cite{Dons2012,Kaur2007}, i.e., in urban, traffic-prone areas spatial variability is very high\cite{Peters2014,Peters2012,Setton2011}.  Several pollution sources have been addressed with success. However, persistent problems remain in urban areas, where traffic and domestic heating are important sources\cite{EuropeanEnvironmentAgency2013}. Next to the technical solutions (e.g., electrical mobility), people's personal perceptions, behavior and choices play a major role in addressing these issues and facilitating change in a bottom-up manner.

Participatory sensing, involving citizens in environmental monitoring, can have multiple potential benefits. Firstly, it can increase coverage of monitored areas, both in time and space, due to the ability to distribute the monitoring activities to multiple individuals\cite{Hasenfratz2012}. Secondly, the act of monitoring pollution by citizens could facilitate learning and increase their awareness of environmental issues\cite{Rio1992}. 
A recent report on environmental citizen science concludes that few studies on public participation in science and environmental education have rigorously assessed changes in attitudes towards science and the environment, and environmental behaviors. There appear to be relatively few examples of participatory citizen science having a tangible impact on decision making, although the potential is often noted\cite{ScienceCommunicationUnit-UniversityoftheWestofEngland-Bristol2013}.

One element to foster large scale participation in participatory monitoring campaigns is the availability of low-cost wearable sensing devices. These will give intrinsically lower quality data, so the trade-off is between the social benefits and the quality of the data\cite{Cuff2008}. 
Several efforts have been made to develop such low-cost wearable sensing devices, integrating low-cost gas sensors, GPS and mobile phones. 
The CommonSense project\cite{Dutta2009} built hand-held devices containing CO, NOx and ozone sensors. Another example, which was quite successful in raising funds through crowdfunding, is the Air Quality Egg\cite{aqe}, designed for static measurements and containing NO$_2$ and CO sensors.

However, many of these projects focus mainly on the electronics and systems integration, power issues, wireless data transfer, data storage and visualization and pay little attention to the limitations and quality issues of the gas sensors adopted. Very few tests or validation results have been published in publicly available reports or peer reviewed literature. Examples are Hasenfratz et al. and Mead et al.. Hasenfratz et al.\cite{Hasenfratz2012} introduce GasMobile, a platform measuring ozone concentration, which is connected to a smartphone by USB. They take into account important issues such as sensor quality, calibration, and effect of mobility on sensor readings. Mead et al.\cite{Mead2013} developed sensor boxes with electrochemical sensors, which entailed changes in the sensor technology itself, in the electronics and complex data analysis. The CitiSense\cite{castell2014} project is currently building an infrastructure for citizen engagement in environmental monitoring. 

Another issue is the collection of  a representative data set using mobile air quality sensing technologies. To be representative and useful for personal or community decision making, mobile measurements have to be repeated regularly, data have to be aggregated over relevant time frames and locations, and carefully interpreted using data handling and expert knowledge to filter out inaccuracies\cite{Peters2012, VandenBossche2015}.The supplementary material  S1 discusses the challenges involved in using low-cost sensors for air quality monitoring and describes the approach used by our project to address quality issues.

An important issue concerns the technological versus social aspect of such projects.  Most of the existing projects concentrate mainly on the sensor side of participatory air quality sensing, i.e., how to build the sensing devices and map pollution. However, participant engagement, participatory patterns, learning and awareness are equally important aspects, and feed back into the quality of the data collection, as we have also shown in a parallel project concerned with noise pollution\cite{Becker2013}. By collecting subjective data as well, monitoring campaigns can enable not only  air quality data collection, but also analysis of volunteer behavior, strategies and a possible increase in awareness. 

In this paper, we discuss the behavior and perceptions of citizens involved in monitoring, during a large scale international test case: the AirProbe International Challenge (APIC)\cite{APIC}. This was organized simultaneously in four cities: Antwerp (Belgium), Kassel (Germany), London (UK) and Turin (Italy). In this test case a web-based game, air quality sensing devices and a competition-based incentive scheme were combined to collect both objective air quality data and data on perceived air quality, to analyze participation patterns and (changes in) perception and behavior of the participants. The test case was organized as a competition between the cities, to enhance participation.  
For the first time to our knowledge, an end-to-end scientific platform for participatory air pollution sensing, developed as part of the EveryAware project\cite{EA}, was used. This platform is described briefly in the Methods section, with more details included in the supplementary material  S1. The quality and representativeness of the collected air quality data are also discussed in  S1.

During this test case, volunteer participants were asked to get involved in two activity types. The first one consisted in using a sensing device (Sensor Box), to measure air pollution (black carbon (BC) concentrations) in their daily life, generating what we call \emph{objective} data. The second activity was playing a web game (AirProbe), where volunteers were asked to estimate the pollution level in their cities by placing flags (so called \emph{AirPins}) on a map and tagging them with estimated black carbon (BC) concentrations on a scale from $0$ to $10$\,\unit{\mu g/m^3}, resulting in \emph{subjective} data on air pollution (perception). Volunteers involved in the measuring activities were encouraged to play the game and bring other players as well (create a team).

The two data types allow for an analysis of user behavior and perception throughout the challenge. To enable this, the test case was composed of three phases. In phase 1, only the online game was available, so we could obtain an initial map of the perceived air pollution. In phase 2 the measurements started in a predefined area in each of the cities (corresponding also to the web game area), with the web game running in parallel. Phase 3 introduced a change in the game, so that players could acquire limited information about the real pollution in their cities in the form of sensor box measurements averaged over small areas (so called \emph{AirSquares}).  At the same time, measurements were continued, this time without a restriction of the area to be mapped. Incentives in the form of prizes were given at the end of each phase to the best teams/players (please see Methods and Supplementary Material  S1 for more details).

 The data collected during the test case are used here to analyze participation patterns, in terms of activity and coverage, and any changes in perception. Our results indicate that better coverage is obtained when volunteers are assigned a specific mapping area, compared to when they are asked to select the time and location of their measurements. Additionally, when allowed to measure freely, they seem to be attracted to places with higher pollution levels.  Furthermore, while at the beginning of the challenge the general perception was that pollution was higher than in reality, perceptions changed in time indicating increased knowledge of real pollution levels. The amount of data collected in the test case, together with the first insights we obtained from it, suggest that bottom-up participatory sensing approaches are effective in attracting participants with high levels of activity and also in enhancing citizen awareness of real pollution levels.


\section*{Results }

Volunteer involvement and activity levels are among the most important elements in participatory monitoring campaigns, since these can determine the success of the campaign. Large activity is required for acquiring meaningful data, both objective, for analysis of the environment itself, and subjective, for analysis of social behavior. 
The test case presented here has successfully involved 39 teams of volunteers in 4 European locations, gathering 6,615,409 valid geo-localized data points during the challenge (the measuring device collects one data point per second). An additional 3,326,956 data points were uploaded to our servers in the same period, but were missing complete GPS information, and were not included in the analysis. Some of these measurements contained labels (tags), with 742 geo-localized overall tags coming mostly from one location of the challenge (London). 
 
 Additional information on perception of pollution has been extracted from the online game. The platform had 288 users in total, over six weeks, 97 of which played the game at least ten times. Their activity resulted in 70,758 AirPins at the end of the test case, which we will use to assess perceived pollution levels.

\begin{figure}
\centering
\includegraphics[width=0.9\textwidth]{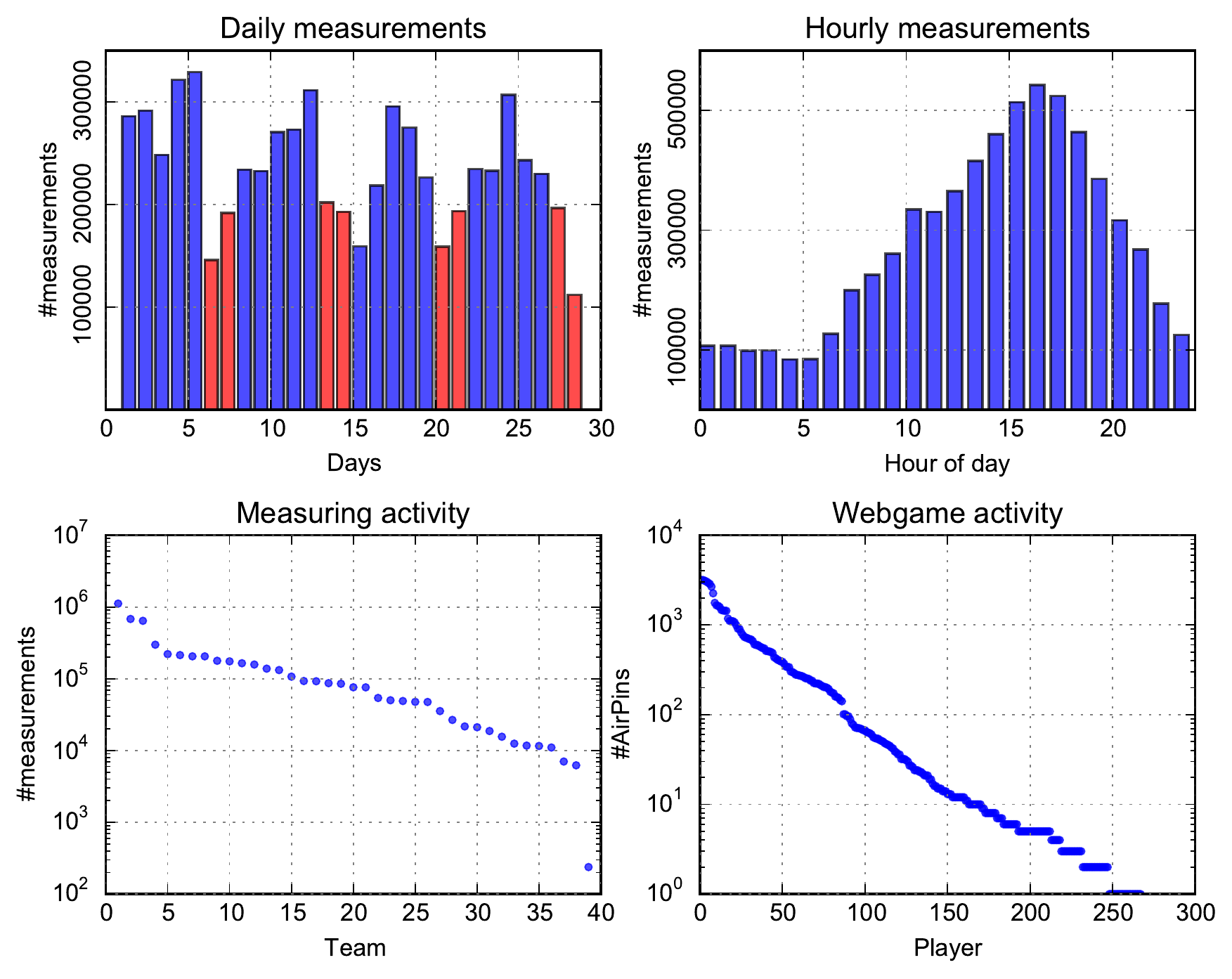}
\caption{{\bf Volunteer activity patterns}. The subplots in the top row show daily (weekends shown in red) and hourly measurements by volunteers. The distribution of the web game activity among players is shown in the bottom-right subplot, while the distribution of the number of measurements performed per team is depicted at the bottom-left (the distributions are displayed by ranking the volunteers by activity and then displaying the number of measurements/AirPins in descending order, using a rank-frequency plot) .
\label{fig:activity}}
\end{figure}

 Figure~\ref{fig:activity} shows general participation patterns, both for the measuring activity and for the web game. Further details about participation, for each of the four locations of the test case, can be found in supplementary file  S1. The daily number of measurements
show larger activity during the week compared to weekends, with almost twice the activity in the peak days (Wednesday/Friday). This indicates that the volunteers were strongly interested in monitoring their exposure in relation to the routine activities of the week, which probably include commuting and access to highly polluted environments. It might also mean that it was easier for participants to monitor as part of their weekly routine whereby at the weekend monitoring would require more effort as it would not comprise part of their commute, for example, or may have impacted on other leisure activities that they wanted to carry out. Daily patterns (hourly measurements) indicate a peak in activity in the afternoon, around 5 pm, again probably due to afternoon commuting. However, measurements are performed at all hours of the day, indicating the presence of very dedicated volunteers. In fact, the total number of measurements per team indicates several teams with very high activity levels, with the most active team reaching almost 1 million points (equivalent to over 270 hours of measurements). However, team activity was very heterogeneous, with some teams collecting much less data than the others. This heterogeneity was found within the same city (e.g., the highly active teams are spread over three of the four cities), indicating that differences in activity were in general based on personal predisposition and not location. However, some of the heterogeneity between the cities can also be explained by the differences in instructions, emphasis and incentives.

The web game activity follows similar heterogeneous patterns. Figure \ref{fig:activity} also shows the distribution of the number of AirPins used to declare perceived pollution levels by game players. Some of them got very involved in this activity, with over 2000 AirPins used, while many players had very low activity (started the game but did not continue). The distributions appear to follow a power law, also typical for other social activity patterns\cite{Garas2012,dynamics2013}. It is important to mention that managing hundreds of AirPins required a large amount of time to be spent in the game, indicating the high involvement levels that the players reached. 


Besides activity in terms of number of measurements, another important aspect is \emph{coverage}, both in \emph{space} and \emph{time}. As we have seen before, measurements have been performed at all hours of the day and days of the week. However, usually not all areas are covered equally. Here we show general information about overall coverage achieved (with more details for each location included in the supplementary file  S1). 

\begin{figure}
\centering
\includegraphics[width=0.9\textwidth]{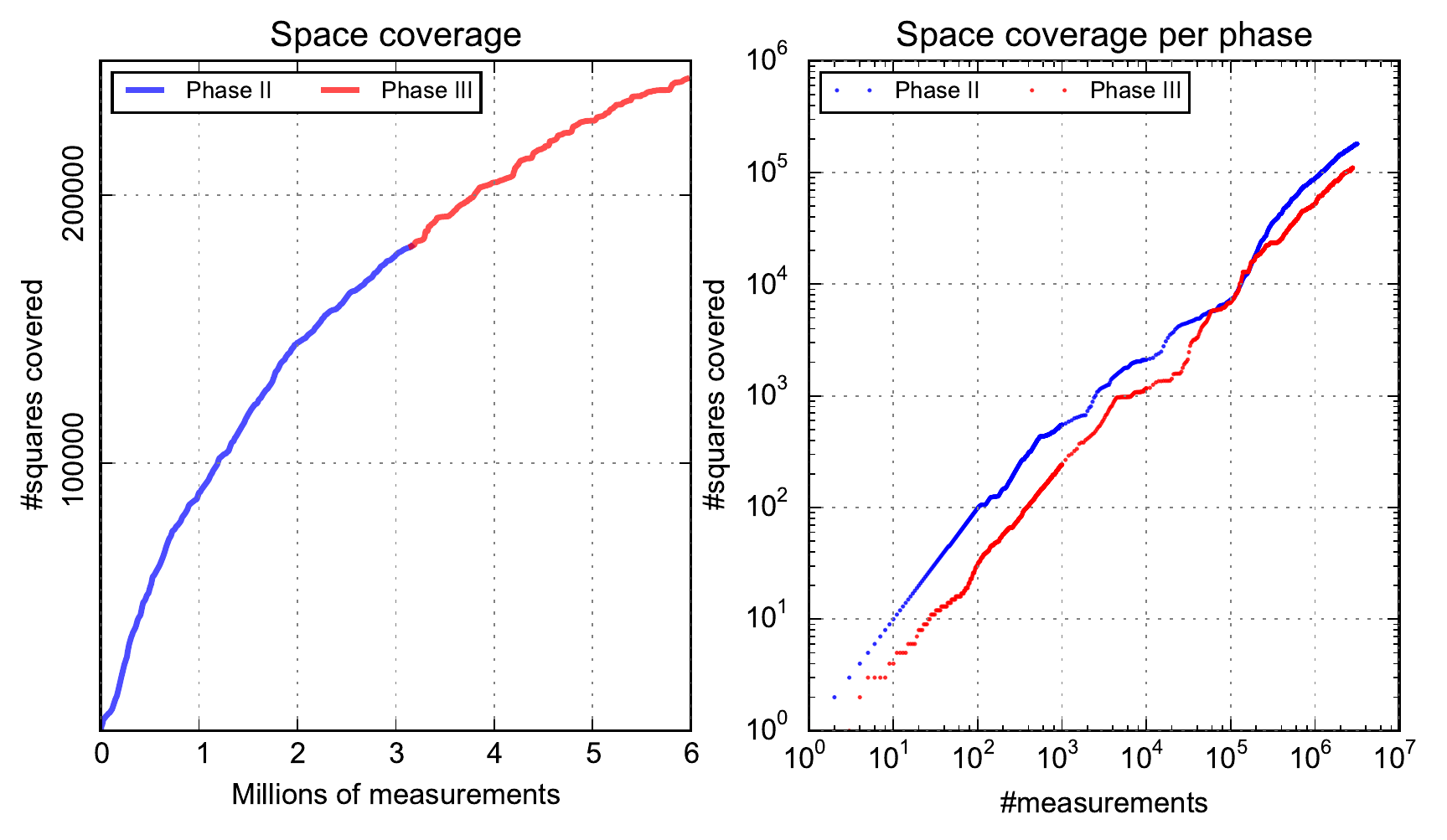}
\caption{{\bf General space coverage data.}  Left panel: growth of the number of squares covered for the entire challenge. Right panel: growth of the number of squares covered per phase, in a log - log plot. 
\label{fig:spacecoverage}}
\end{figure}

In order to compute the coverage, the area of each of the four participating cities was divided into 10 by 10 meter squares (tiles). One square was considered covered if at least one measurement was performed within it. Figure \ref{fig:spacecoverage} shows how the number of squares covered grows as users perform more measurements, both overall and for each phase individually. The volunteers had different tasks in the two measuring phases (phase 2 and 3 of the test case). In phase 2, they had to concentrate on covering as much as possible of a specific area, while in phase 3 they could explore any area they wanted. 

\begin{figure}
\centering
\includegraphics[width=0.9\textwidth]{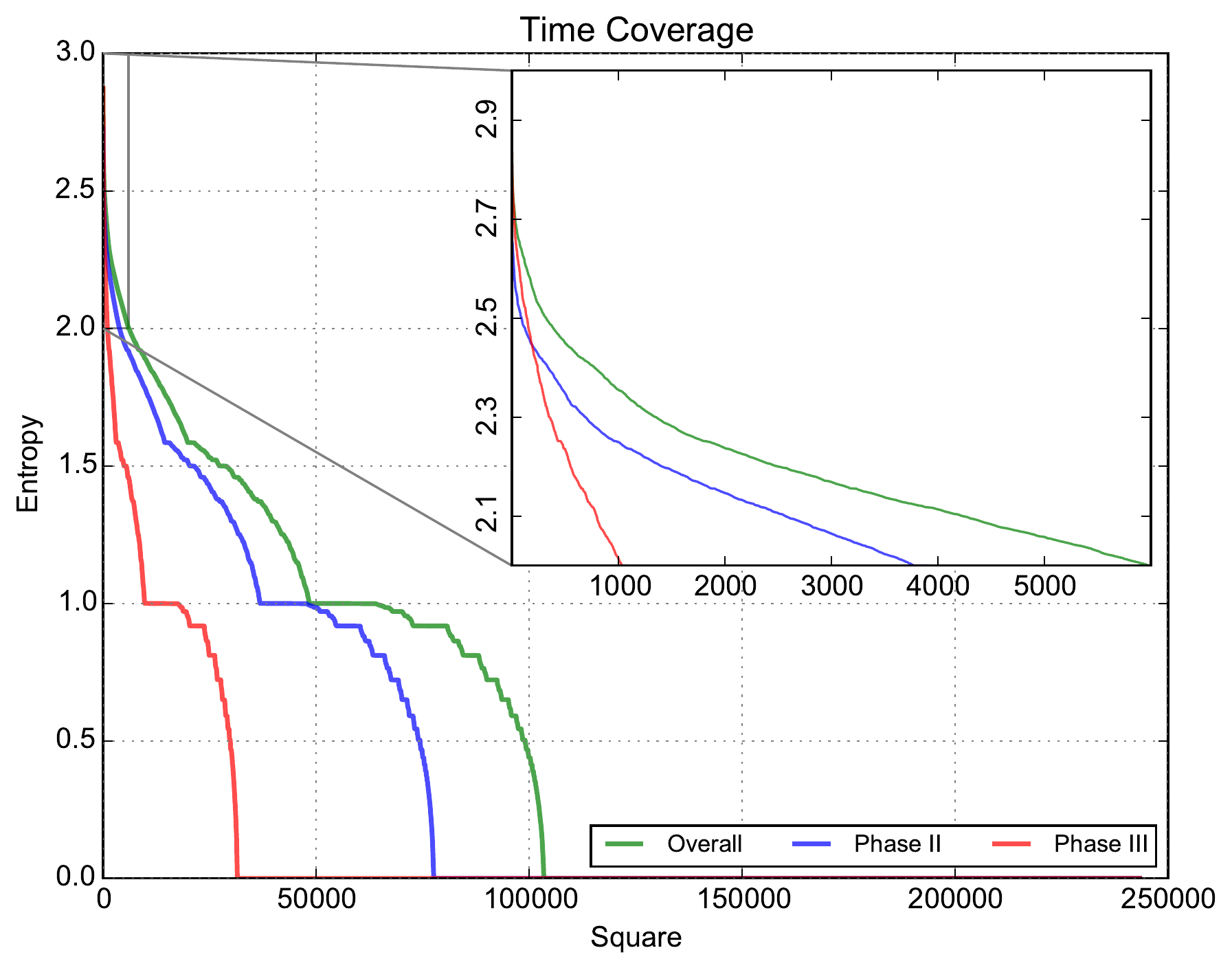}
\caption{{\bf General time coverage data.} Time coverage per phase and overall. The inset shows an enlarged view of the leftmost part of the plot (top ranked squares). 
\label{fig:timecoverage}}
\end{figure}

Figure \ref{fig:spacecoverage} indicates that space coverage grows steadily with the number of measurements, meaning that users continue to explore new areas over the course of the challenge. However, while at the beginning of the challenge the growth is fast, this decreases in time. This indicates less exploration as the challenge evolves, due to the fact that volunteers measure at the same location multiple times. When looking at individual phases, it appears that during phase 2 space coverage was much better than in phase 3. This does indeed mean that volunteers displayed a better exploratory behavior at the beginning and when asked to cover a specific area of the city, compared to when they were asked to map any place they wished. In the latter case, they went for their daily routes that were not so extensive, and did not explore further. For both phases the growth of the space coverage follows a power-law, with exponent 0.73 in phase 2 and 0.79 in phase 3. This suggests that, although on the short term, space coverage in phase two is larger, in the long run  the strategy of phase 3 might actually produce better coverage. However, the restricted time frame of our challenge can not provide further proof for this hypothesis.
Since pollution levels vary both in time and space, it is important to have more measurements in the same location. So, for each tile, we also look at how measurements are spread in time, i.e., time coverage. We divided the measurements into 8 categories based on the time of measurement: 4 working day categories and 4 weekend categories, with time thresholds at hours 08:00, 14:00, 18:00 and 23:00. Measurements on Friday after 23:00 fall in the working day category, while those on Saturday before 08:00 in the weekend category. The entropy of the resulting sets was computed. For each square, we obtained the fraction $f_i$ of measurements in each category $i$ as the ratio between measurements falling into that category and the overall number of measurements in that square. Then the entropy for that square is $S=-\sum_{i=1}^8{f_i \log_2 f_i}$. A higher entropy indicates a better spread of measurements in time. 
Figure \ref{fig:timecoverage} shows the distribution of the entropy for all squares covered, in a rank-entropy plot (squares are sorted descending by entropy and the entropy values plotted for each square). A few squares had a very good time coverage and they  correspond, most likely, to hubs in the four cities (e.g., popular leisure locations or transportation hubs). At the other extreme there are many squares (more than half) that have been covered only in one time slot (entropy is 0). Between the two extremes, time coverage is dropping fast when moving through the ranked squares.

 The curves display jumps and it appears that  squares can be divided into sets based on time coverage. One first set (rightmost) includes those squares that have measurements only at one time of the day (entropy 0), which is followed by those covered in 2 time slots, ending with those that are covered at all times of the day (leftmost). Within each set, coverage decays differently. While for the highly covered squares decay appears to be exponential (as plotted in the inset), this becomes slower as the coverage decreases, with curves resembling polynomial decay.  

When comparing the two phases, time coverage in phase 2 is much better overall than in phase 3. This indicates that volunteers not only explored more in space, but also in time, during phase 2, while in phase 3 they followed their daily schedule which allowed for poor time coverage as well. This underlines again the importance of giving volunteers a specific mapping area in order to obtain better measurement spread. 

\begin{figure}
\centering
\includegraphics[width=0.9\textwidth]{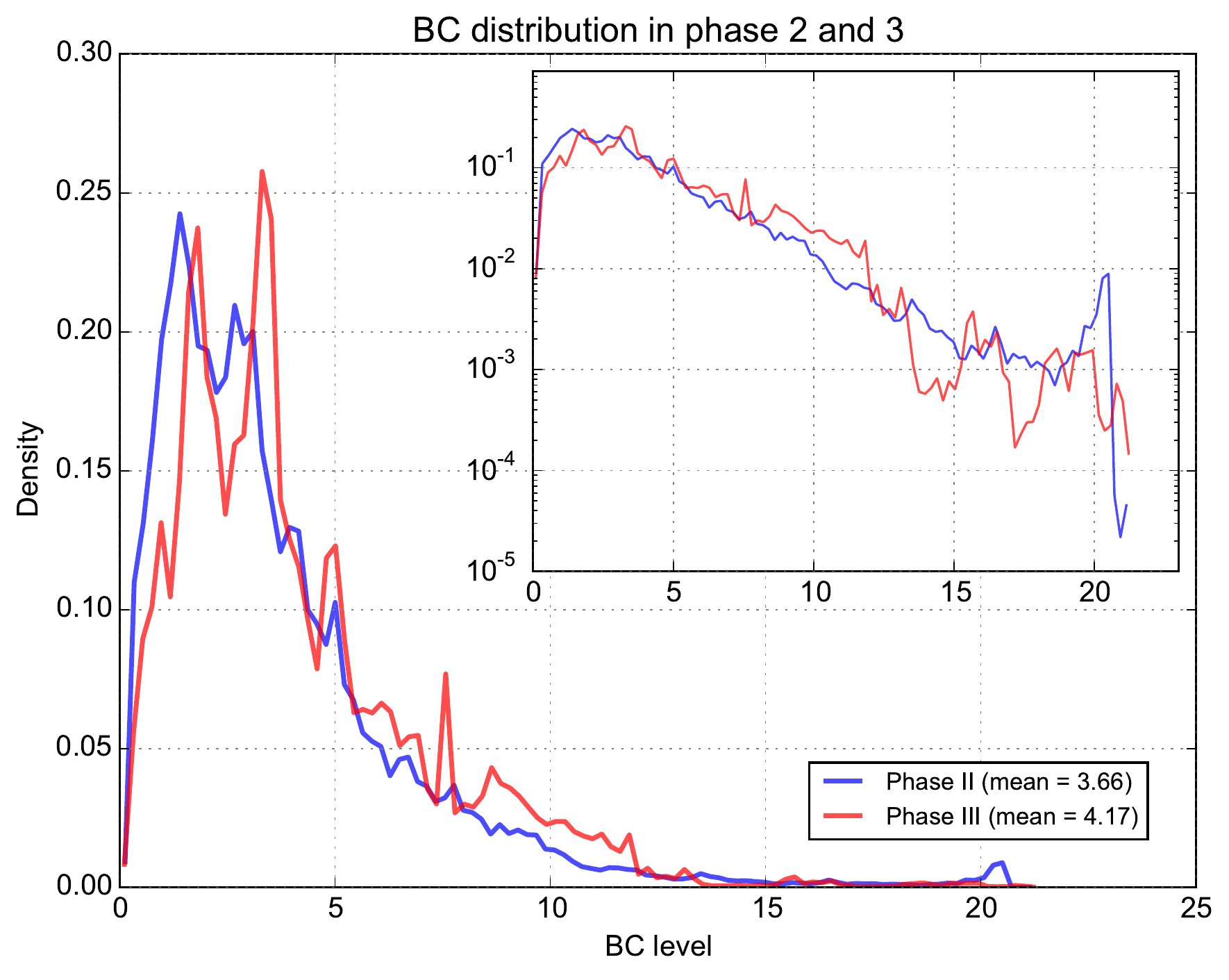}
\caption{{\bf Overall pollution levels compared between the two phases.} The distribution of BC levels are shown for the two measuring phases of the challenge. The inset shows the same plot but with a logarithmic vertical axis, to emphasize the tail of the distribution. 
\label{fig:p1vsp2o}}
\end{figure}

The measured BC levels can also provide useful insight into the aims and strategies of the volunteers during the challenge. To this end, we can examine how these change from phase 2 to phase 3. Thus, Figure \ref{fig:p1vsp2o} shows graphs of BC levels measured in the two phases, and we can observe larger BC values in phase 3 (the distribution is shifted to the right). A Kolmogorov-Smirnov test was performed to test whether differences are significant and a p-value of 2.2e-16 was obtained, confirming the difference. When volunteers can freely choose where to take measurements, it appears that they primarily target more polluted areas. 
When the mapping area is restricted, they tend to have a more systematic approach and cover lower pollution levels as well.
One may argue that pollution levels may change naturally from one day to another, so the shift we see could be do to a higher average pollution level from phase 2 to phase 3. However, comparison with reference data seem to suggest that this is not the case (supplementary material  S1). Additional comparisons per location are also included in S1.

\begin{figure}
\centering
\includegraphics[width=0.9\textwidth]{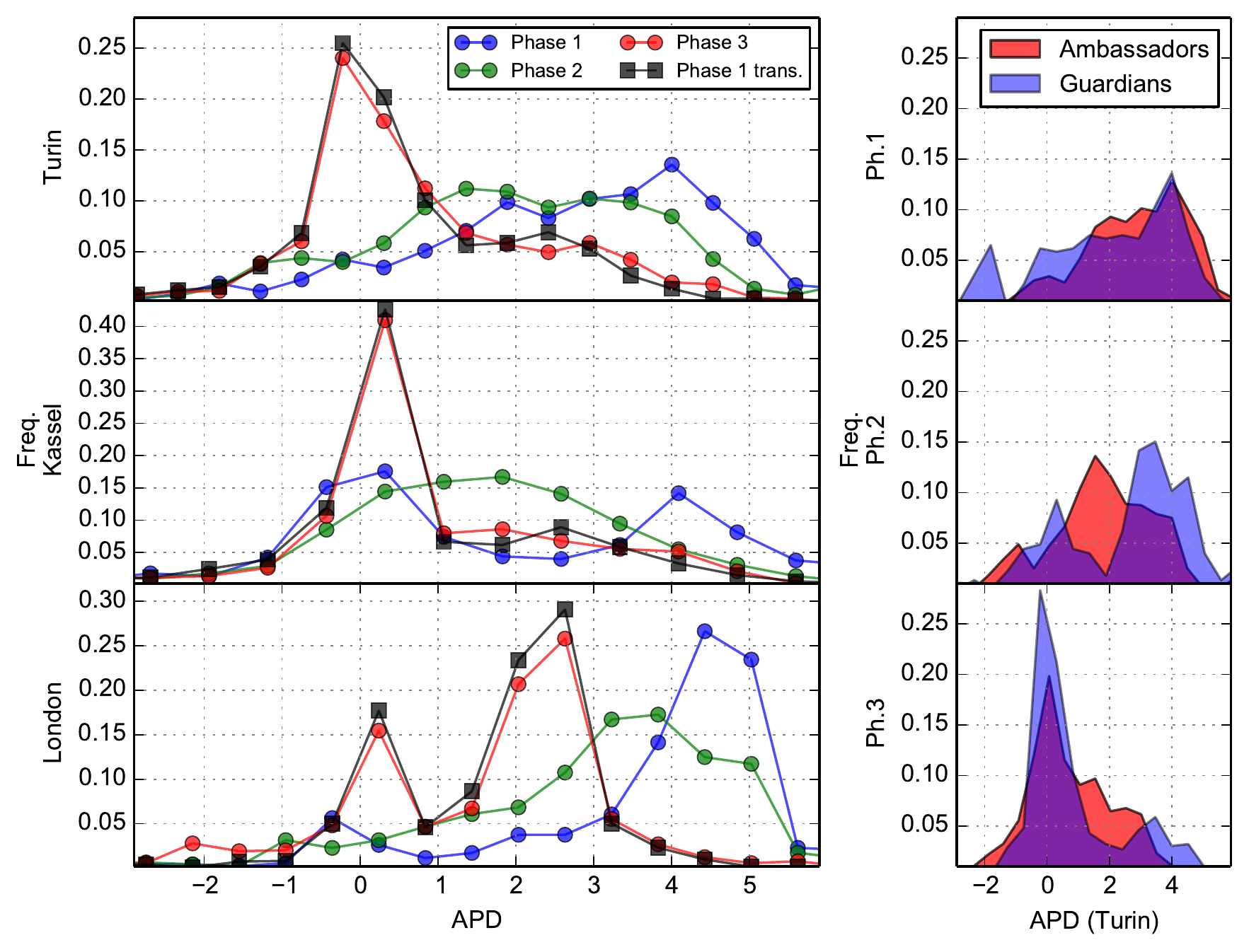}
\caption{{\bf Web game data: APD distributions.} APD is the deviation between the air quality level annotated (the AirPin value) and the aggregated measurements from sensor boxes (the AirSquare).
The left part shows the distributions of the deviations in each phase for Turin, Kassel and London. An estimation of phase 3 distribution elaborated from phase 1 data with our model is also shown (Phase 1 trans.).
The right part shows the distributions for Turin in each phase for AirAmbassadors (volunteers with sensor box that played the web game) and AirGuardians (only web game players).
\label{fig:game}}
\end{figure}

The analysis of the structure and location of the collected objective data gives some insight into volunteer behavior and interests when measuring air pollution. Subjective data, on the other hand, can provide a stronger indication of changes in perception. For this, we look at the data collected by the web game, which consists of perceived levels of pollution in the mapping area, the AirPin values. In particular, to inspect awareness improvement and the learning process, we are interested in the relation between these annotations and the `true' pollution values available in the web game during phase 3 in the form of AirSquares. Thus we define the APD (AirPin difference) as the difference between the AirPin value (perception of the volunteer) and the relative AirSquare value (real pollution level). In other words, the APD is the amount of `error' in the annotation intended as distance from the measurement.
Figure~\ref{fig:game} shows several distributions of the APD. In the left part we have APD distributions in each phase for Turin, Kassel and London. Antwerp did not reach the critical mass of data required for this analysis (the number of web game volunteers was very restricted). 

In phase 1, when no volunteer had been exposed to real measurements, we observe three different opinion structures in the three cities, representing the initial perception of volunteers. A systematic overestimation of pollution is present, i.e., the APD has peaks at $\sim4$\,\unit{\mu g/m^3}. This is likely to be caused by a scale misunderstanding: players, which were not accustomed to the BC concentration scale, almost ignored completely which values were to be considered reasonable and thus used the middle of the scale (i.e., $5$\,\unit{\mu g/m^3}) as a 'normal' value. This results in the observed overestimation since the real average BC concentration measured lies between $1$ and $2$\,\unit{\mu g/m^3}. 

In phase 2 things began to change. Some volunteers (so called \emph{Air Ambassadors}) were given the sensor boxes to start performing measurements. The web game players consisted of these volunteers plus a set of other players recruited by them (so called \emph{Air Guardians}). No data, except for the direct feedback from the boxes, was shown to the volunteers. Even so, a change is visible in the distribution of APD reported in the left part of Figure~\ref{fig:game}. By observing the measurements from their sensor boxes, Volunteers learn that in general BC concentrations are lower than what they believed, and respond by changing the values of the AirPins or taking the information into account when placing new ones. Since the change is quite significant, we also believe that those volunteers with the sensor boxes spread the information about what they were measuring, so that all players changed their perception. This decrease in the pollution levels reported in the subjective data of phase 2 is a first strong indication of learning during this phase. The right side of Figure~\ref{fig:game} shows APD distributions separately for AirAmbassadors (performing measurements) and AirGuardians (who had no direct exposure to measurements until phase 3). We analyzed just the Turin dataset because in the other cities there was no clear distinction due to Ambassadors sharing their sensor boxes. The opinion shift in phase 2 is very strong  for AirAmbassadors, but some change is also visible for AirGuardians, at least for part of the AirPins. This indicates that there was interaction among players, so that not only volunteers performing measurements, but some of their friends also, changed their perceptions. 

Phase 3 brought an important change in the web game. AirSquares were made available, so players could acquire aggregated information (punctual information would have been just copied by the users) in form of average pollution levels within the respective square measured by the sensor boxes. There is a corresponding radical change in the subjective air pollution estimation emerging clearly in the left part of Figure~\ref{fig:game}. In all cities, there is a peak around zero in phase 3 in the APD distribution, meaning there were more players estimating the air quality correcly. This was in some way expected, since we are giving strong hints about pollution levels by means of AirSquares, but there is something more happening. In London there is another bigger peak and also in the other cities the distributions show some asymmetry, pointing out that people are not trusting the hints completely because in that case the distribution would have been more similar to a delta function, i.e., narrow and symmetric. 

In order to describe this phenomenon we defined a stochastic transformation to reproduce the APD distribution for phase 3 starting from the APD distribution of phase 1. This transformation should reproduce the effect of the hints received by our volunteers on the initial distribution of their errors. Based on the empiric observation, the transformation takes into account two main effects: the possibility of complete trust in the hint, so that the opinion is reset near the hint, and the possibility of incomplete trust, so that the opinion is just shifted closer to the hint. The mathematical definition can be found in the supplementary material ( S1).
The left part of Figure~\ref{fig:game} shows, for each location, how the transformed phase 1 data (black squares) matches phase 3 distributions, and this has also been confirmed with statistical procedures described in Methods and in the supplementary material  S1. This provides an indirect proof of the assumptions of our model on the effect of objective data (complete and incomplete trust). Also, we were able to measure the `trust' in the hints for the three cities, by fitting the model to data. We obtained the lowest trust values in London and the highest ones in Turin (full results are reported in the supplementary material  S1).


\section*{Discussion}

Volunteer participation is crucial for the success of bottom-up monitoring campaigns, however most projects concerned with air pollution monitoring concentrate only on the development of the technical tools necessary. Here, we give a different user-centric perspective, using the experience from the EveryAware project, through its large scale international challenge, APIC. The tools developed by the project are described in more detail in the supplementary material  S1. During the challenge both objective and subjective data were collected, and used here to analyze participatory patterns and possible changes in behavior or perception. 

Objective measurements allowed for analysis of user interests during the challenge and activity patterns. A large number of measurements was obtained, however, coverage varied from location to location, with higher values when monitoring areas were restricted. Both coverage and pollution levels measured indicated a volunteer tendency to monitor familiar areas when there was no restriction, with a search for highly polluted spots. 

Subjective data, on the other hand, allowed for analysis of perceived pollution levels and learning mechanisms. 
We observed, by analyzing differences between perceived and real pollution levels, that users are able to reduce the `errors' in the annotations, by learning the true values. However, some inertia in changing the old opinion structure was also observed, since asymmetric tails and slow shifts of old peaks are present. We also looked at differences between AirAmbassadors (volunteers with sensor boxes that played the web game) and AirGuardians (only web game players). In phase 1 there is no clear distinction between them, as it is expected. In phase 2 Ambassadors, who begin to learn real pollution levels from the sensor boxes, start to shift their opinions, reducing the errors, while Guardians change less. Finally, in phase 3 we observe Ambassadors continuing to shift their opinions in a smooth way, with a certain inertia, while Guardians change radically showing a prominent primary peak at zero estimation error with a secondary peak in the position of the old peak. We can argue that the personal experience of the Ambassadors produces a smoother transition (which begins in phase 2), while the in-game information produces radical changes. But still both approaches shows the inertia we described earlier, even if in different forms.

In general, we can conclude that all our evidence shows that involving volunteers in monitoring campaigns can result in large amounts of data collected. These data show that participation can help learning, to create a more accurate perception of air quality. Thanks to our case study, it has also been possible to outline some of the mechanisms behind the resistance of subjective opinions to objective results.

 
\section*{Materials and Methods}

The study presented here is based on data collected by volunteers during a large scale test case (AirProbe International Challenge - APIC) organized in four European cities (Antwerp, Kassel, London and Turin) in from October 2013 to November 2013. It required volunteers to measure air quality as well as provide their opinion on air pollution, using the EveryAware platform. This consists of a sensing device (Sensor Box), measuring air pollution, a mobile application (AirProbe), allowing for data visualization and upload to servers, a set of web services and websites, handling data storage and visualization and a web game developed on the XTribe platform\cite{Caminiti2013}, allowing to collect individual perceptions of pollution. In the following we provide a brief description of each of the components and of the tools used for data analysis, with further details included in the supplementary file  S1.

\subsection*{Ethics statement }
This work is part of the European project Every Aware, contract number IST-265432. The European Commission finances only those projects that comply to its ethics and privacy regulations. Citing from the regulations of the Seventh Framework Programme, Decision No 1982/2006/EC, Article 6: ``All the research activities carried out under the Seventh Framework Programme shall be carried out in compliance with fundamental ethical principles."  At the same time, the official rules for participation, Article 15, mention: ``A proposal which contravenes fundamental ethical principles shall not be selected. Such a proposal may be excluded from the evaluation and selection procedures at any time". Hence, acceptance and funding of this work by the European Commission implies approval of the ethics statement made in the proposal. This is why no further formal ethics approval was required for this research to be performed. 

All participants to our study had to participate in training for using the sensor box and install our mobile application. Before admission to the test case, all volunteers were required to sign our Terms and conditions, which represents the user's consent to use the measurements made.  These clearly state that the data will be used for research purposes only and no personal information will be made public or used for other purposes. 

Volunteers were recruited using a range of approaches in each city. These included a designated Facebook page, the EveryAware project website, posters, newspaper articles and either university mailing lists or those of local interests groups and environmental agencies (see the methods section 'Case study' and supplementary material S1 for further details). All volunteers could leave the study at any stage, however none chose to do so. All volunteers named in the Acknowledgements section gave specific permission to be named.

\subsection*{Sensing device: the sensor box }

The sensor box contains a sensor array of 8 commercially available gas sensors and two meteorological sensors (temperature and humidity). The gas sensor array consists of low-cost continuous sensors of CO, NOx, O$_3$ and VOC, which are important pollutants in the urban outdoor environments. These pollutants are either directly emitted by vehicles or other combustion processes, or formed from emitted precursors in the vehicle exhaust. The main criteria for sensor selection were the specific requirements posed by the mobile use of the sensor box for air quality monitoring as well as the hardware compatibility with the box. The gas sensors were examined by a range of performance tests under laboratory and outdoor conditions. These tests showed that none of the individual sensors can be used on its own. The observed selectivity, stability and response times of the different sensors introduced the need for a multivariate calibration procedure for the sensor boxes. Performance tests and calibration are described in more detail in the Supplementary material  S1.        

\begin{figure}
  \centering
 \includegraphics[width=\textwidth]{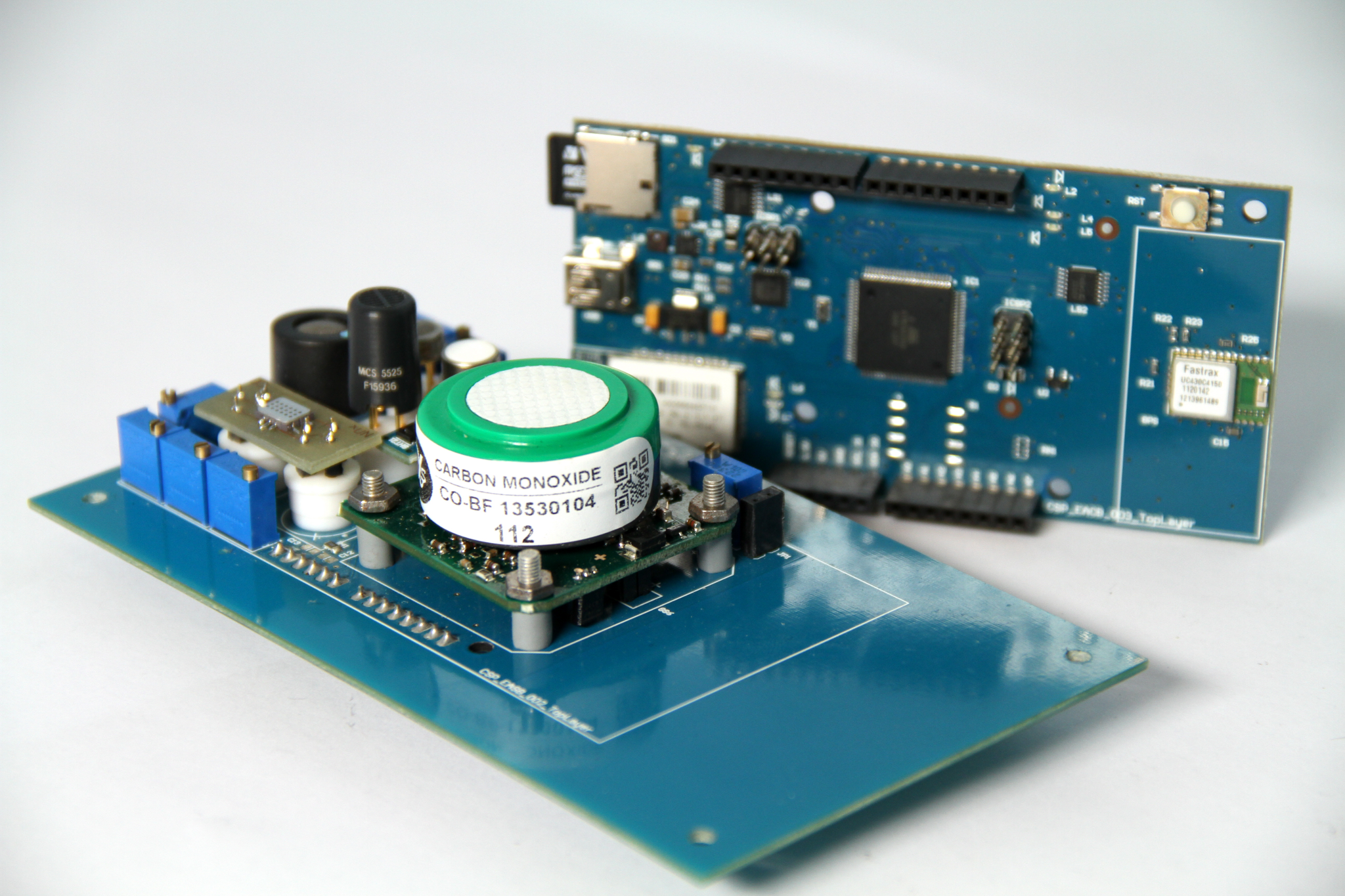}
  \caption{ {\bf Sensing device.} The two electronics boards of the sensor box with the gas sensors mounted on top of the sensor shield.\label{fig_sensorbox}}
\end{figure}

The sensor box electronic system has been designed with the purpose of being a low-cost, open and scalable platform. 
It is composed of two main boards (Fig.\,\ref{fig_sensorbox}). The first is a general purpose one that includes basic storage (micro SD card), positioning (GPS) and communication (Bluetooth) capabilities, while the second is a sensor shield able to host all gas sensors.
The design is based on Arduino components and it is completely open source, so that anyone can reproduce and modify the hardware or even use the original hardware and develop different software to be run on it.

\subsection*{The AirProbe mobile application }

AirProbe is an Android application designed to connect to the sensor box 
via Bluetooth, acquire sensor readings and transit them to the EveryAware servers 
as soon as a working connection to the Internet becomes available.
In addition, the application allows users to visualize the data they collect. Specifically, they can see their tracks on a map,
calculate an estimated black carbon exposure and follow sensor output in real time plots. While collecting data, users can make free annotations (tags) that will be attached to the recordings and sent to the servers.

\subsection*{Web platform }


The case study web platform\cite{becker-2013} is designed for collecting,
storing, retrieving, analysing and visualizing large amounts of data data 
from different data sources.
It provides endpoints for application like the AirProbe mobile application 
to upload data to.
These data are then processed and cleaned, with several statistics and 
visualizations available on a public as well as a personal level.
This facilitates further analysis and deeper understanding of the data by the user.

A collection of statistics pages provides overall information about the data, 
such as graphs showing currently active sensor boxes, the overall black carbon 
average per day, or the overall number of collected measurements per day. 
Also, information on separate sessions corresponding to different tracks 
(defined both by the Sensor Box and by the user) is available.
This allows users to compare routes and locations. 
A world map gives a visual overview on the collected data.
This includes cluster and grid views as well as a heatmap representation of the 
collected data on a personal as well as a global level providing visual information
about areas with good measurement coverage and their average pollution levels.
Users also have the possibility of downloading their own data, in case they want 
to compile any further personal statistics.

During the APIC challenge, the platform was specifically tuned for the needs of the 
game.
Even though the platform supports several statistics and visualization of the data, 
most of this functionality has been disabled during the second stage of the challenge, in order
to make opinions on air quality during the web game as unbiased as possible.
The goal was for the AirAmbassadors and their sensor boxes to be the sole source of
information regarding real measurements in order to limit information flow and facilitate a more controlled environment for the experiment. All visualisations were back online in the third phase of the challenge.

The web platform has been also providing 
a ranking page for the AirAmbassadors to be motivated throughout the challenge.
Points were issued for space and time coverage during each collection phase. 
The ranking page showed which city and which team was ranked first globally as 
well as per city.
In addition, the AirAmbassadors and their teams were able to access several statistics about 
their measurement behavior and the data collection process, 
including a coverage heatmap, the amount of covered squares and their points.

\subsection*{The web game }

The AirProbe web game is a simplified map management game. Players are called to fulfil their role of Air Guardians by annotating the map with so-called \emph{AirPins}: geo-localized flags tagged with an estimated or perceived pollution level (black carbon concentration in \unit{\mu g/m^3}, on a scale from $0$ to $10$).
The game area of each city is divided into tiles. At the beginning of the game, users are asked to create a profile (by choosing an avatar and a name) and to choose a city and a team. Then the volunteer starts from a given tile of the map of the chosen city. Users can interact by placing (or editing or removing) AirPins or by expanding their territory, i.e., buying more tiles. Each day, the AirPins placed generate a revenue based on the precision of the annotation (precision depends on what other users think of the same area). In order to collect the revenue generated every day by each AirPin, the user has to access the game daily, otherwise the revenue will be lost. The collected revenue will be added to the user balance, allowing them to buy more AirPins and more tiles. In this way, players can build their air pollution perception map.  
At the beginning of phase 3, a new feature was made available in the web game: the AirSquare map. This consisted in an alternative map on which players could buy AirSquares, i.e., information about measured pollution levels aggregated on a small area. This data spreading stimulated the learning process described earlier.

\subsection*{Case study }

In order to set up the APIC study, volunteers were recruited in each of the four cities and they
comprised two types of participants: Air Ambassadors, who were tasked with collecting air quality 
measurements with the sensor box, playing the online game, and recruiting Air Guardians, and Air 
Guardians, whose central focus was to play the online game and who were linked to a team of Air 
Ambassadors. Volunteers were recruited using a range of approaches in each city. These included a 
designated Facebook page, the EveryAware project website, posters, newspaper articles and either 
university mailing lists or those of local interests groups and environmental agencies (see supplementary material  S1 for 
further details). 

Incentives were offered during the initial call to participate in the study with the aim to encourage 
participation and maintain engagement. Prizes were given out to the team of Air Ambassadors with 
the best temporal/spatial air quality measurement coverage and the most active Air Guardians in 
each city over the different phases. Various strategies were incorporated into the online game to 
encourage ongoing play and the prizes related to the number of days played and the total revenue
gained for each day of play. The rewards offered varied slightly across the four cities and are detailed 
in the supplement.

\subsection*{Data analysis }

To model the evolution between the phases of the APD distribution represented in the left part of Figure~\ref{fig:game} (Phase 1 trans.), we implemented a simple modeling approach rearranging the opinions depending on their distances from the hint which is defined in the supplementary material  S1. The transformation introduces 4 parameters, quantifying the inertia effects in the opinions shift. 
To check the quality of our model and to determine the values of parameters introduced we used a Kolmogorov-Smirnov test applied to the phase 3 dataset and to the phase 1 transformed dataset. Since it is a stochastic model, we performed several applications and found a convincing result for the $p_{val}$ of  $20\%$, which means that the hypothesis is consistent with observations. More details are provided in the supplementary material  S1.

\section*{Supporting Information}

\subsection*{S1}
\label{S1}
{\bf Platform description and further data analysis.} Details for the different platform components and data features can be found in this file.

\section*{Acknowledgments}
This research has been supported by the EveryAware project funded by
the Future and Emerging Technologies program (IST-FET) of the European
Commission under the EU RD contract IST-265432. The  framework in which this study was developed was built and maintained by the EveryAware consortium with teams coming from several institutions across Europe, as listed in the author affiliations. We would like to thank all the volunteers at the four locations for participating in our test cases and enabling this research .


\bibliographystyle{plain}

\bibliography{airprobe}

\begin{thebibliography}{10}

\bibitem{Rio1992}
UN~General Assembly.
\newblock {Rio declaration on environment and development}.
\newblock {\em Agenda}, 21, 1992.

\bibitem{Becker2013}
Martin Becker, Saverio Caminiti, Donato Fiorella, Louise Francis, Pietro
  Gravino, Mordechai~(Muki) Haklay, Andreas Hotho, Vittorio Loreto, Juergen
  Mueller, Ferdinando Ricchiuti, Vito D.~P. Servedio, Alina S\^irbu, and
  Francesca Tria.
\newblock Awareness and learning in participatory noise sensing.
\newblock {\em PLoS ONE}, 8(12):e81638, 12 2013.

\bibitem{becker-2013}
Martin Becker, Juergen Mueller, Andreas Hotho, and Gerd Stumme.
\newblock A generic platform for ubiquitous and subjective data.
\newblock In {\em 2013 ACM International Joint Conference on Pervasive and
  Ubiquitous Computing , UbiComp 2013; 1st International Workshop on Pervasive
  Urban Crowdsensing Architecture and Applications, PUCAA 2013, Zurich,
  Switzerland -- September 8-12, 2013. Proceedings}, pages 1175--1182, New
  York, NY, USA, 2013. ACM.

\bibitem{Caminiti2013}
S.~Caminiti, C.~Cicali, P.~Gravino, V.~Loreto, V.D.P. Servedio, A.~Sirbu, and
  F.~Tria.
\newblock Xtribe: A web-based social computation platform.
\newblock In {\em Cloud and Green Computing (CGC), 2013 Third International
  Conference on}, pages 397--403, Sept 2013.

\bibitem{castell2014}
N.~Castell, M.~Kobernus, H.~Liu, P.~Schneider, W.~Lahoz, A.~J. Berre, and
  J.~Noll.
\newblock {Mobile technologies and services for environmental monitoring: The
  Citi Sense MOB approach.}
\newblock {\em Urban Climate}, 2014.

\bibitem{Cuff2008}
Dana Cuff, Mark Hansen, and Jerry Kang.
\newblock Urban sensing: out of the woods.
\newblock {\em Commun. ACM}, 51(3):24--33, 2008.

\bibitem{Dons2012}
Evi Dons, Luc {Int Panis}, Martine {Van Poppel}, Jan Theunis, and Geert Wets.
\newblock {Personal exposure to Black Carbon in transport microenvironments}.
\newblock {\em Atmos. Environ.}, 55(0):392--398, August 2012.

\bibitem{Dutta2009}
Prabal Dutta, Paul~M. Aoki, Neil Kumar, Alan Mainwaring, Chris Myers, Wesley
  Willett, and Allison Woodruff.
\newblock Common sense: Participatory urban sensing using a network of handheld
  air quality monitors.
\newblock In {\em Proceedings of the 7th ACM Conference on Embedded Networked
  Sensor Systems}, SenSys '09, pages 349--350, New York, NY, USA, 2009. ACM.

\bibitem{EuropeanEnvironmentAgency2013}
{European Environment Agency}.
\newblock {Air quality in Europe - 2013 report}.
\newblock Technical report, 2013.

\bibitem{EA}
{EveryAware Consortium}.
\newblock { EveryAware Project, www.everyaware.eu}, 2012 Date of access:
  03/02/2015.

\bibitem{APIC}
{Everyaware Consortium}.
\newblock { AirProbe International Challenge -
  www.everyaware.eu/category/apic/}, 2013 Date of access: 03/02/2015.

\bibitem{Garas2012}
Antonios Garas, David Garcia, Marcin Skowron, and Frank Schweitzer.
\newblock {Emotional persistence in online chatting communities}.
\newblock {\em Nat. Sci. Rep.}, 402:1--34, 2012.

\bibitem{Hasenfratz2012}
David Hasenfratz, Olga Saukh, Silvan Sturzenegger, and Lothar Thiele.
\newblock Participatory air pollution monitoring using smartphones.
\newblock {\em Mob. Sens.}, 2012.

\bibitem{Kaur2007}
S~Kaur, M~J Nieuwenhuijsen, and R~N Colvile.
\newblock {Fine particulate matter and carbon monoxide exposure concentrations
  in urban street transport microenvironments}.
\newblock {\em Atmos. Environ.}, 41(23):4781--4810, 2007.

\bibitem{Lave2013}
Lester~B Lave and Eugene~P Seskin.
\newblock {\em Air pollution and human health}, volume~6.
\newblock Routledge, 2013.

\bibitem{Mead2013}
M.I. Mead, O.A.M. Popoola, G.B. Stewart, P~Landshoff, M~Calleja, M~Hayes, J.J.
  Baldovi, M.W. McLeod, T.F. Hodgson, J~Dicks, A~Lewis, J~Cohen, R~Baron, J.R.
  Saffell, and R.L. Jones.
\newblock {The use of electrochemical sensors for monitoring urban air quality
  in low-cost, high-density networks}.
\newblock {\em Atmos. Environ.}, 70(0):186--203, May 2013.

\bibitem{Peters2012}
J~Peters, J~Theunis, M~{Van Poppel}, and P~Berghmans.
\newblock {Monitoring PM10 and ultrafine particles in urban environments using
  mobile measurements}.
\newblock {\em Aerosol Air Quality Res.}, 13:509--522, 2013.

\bibitem{Peters2014}
J~Peters, J~{Van Den Bossche}, M~Reggente, M~{Van Poppel}, B~De~Baets, and
  J~Theunis.
\newblock {Cyclist exposure to UFP and BC on urban routes in Antwerp, Belgium}.
\newblock {\em Atmos. Environ.}, 92:31--43, 2014.

\bibitem{Lancet2013}
O~Raaschou-Nielsen, ZJ~Andersen, R~Beelen, E~Samoli, M~Stafoggia, G~Weinmayr,
  B~Hoffmann, P~Fischer, MJ~Nieuwenhuijsen, B~Brunekreef, WW~Xun,
  K~Katsouyanni, K~Dimakopoulou, J~Sommar, B~Forsberg, L~Modig, A~Oudin,
  B~Oftedal, PE~Schwarze, P~Nafstad, Faire~U De, NL~Pedersen, C-G Ostenson,
  L~Fratiglioni, J~Penell, M~Korek, G~Pershagen, KT~Eriksen, M~Sorensen,
  A~Tjonneland, T~Ellermann, M~Eeftens, PH~Peeters, K~Meliefste, M~Wang,
  B~Bueno-de Mesquita, TJ~Key, Hoogh~K de, H~Concin, G~Nagel, A~Vilier,
  S~Grioni, V~Krogh, M-Y Tsai, F~Ricceri, C~Sacerdote, C~Galassi, E~Migliore,
  A~Ranzi, G~Cesaroni, C~Badaloni, F~Forastiere, I~Tamayo, P~Amiano,
  M~Dorronsoro, A~Trichopoulou, C~Bamia, P~Vineis, and G~Hoek.
\newblock Air pollution and lung cancer incidence in 17 {E}uropean cohorts:
  prospective analyses from the {E}uropean {S}tudy of {C}ohorts for {A}ir
  {P}ollution {E}ffects ({ESCAPE}).
\newblock {\em Lancet Oncol.}, 14:813--822, 2013.

\bibitem{ScienceCommunicationUnit-UniversityoftheWestofEngland-Bristol2013}
{Science Communication Unit - University of the West of England - Bristol}.
\newblock {Science for Environment Policy In-Depth report : Environmental
  Citizen Science}.
\newblock Technical report, 2013.

\bibitem{Setton2011}
E~Setton, J~D Marshall, M~Brauer, K~R Lundquist, P~Hystad, P~Keller, and
  D~Cloutier-Fisher.
\newblock {The impact of daily mobility on exposure to traffic-related air
  pollution and health effect estimates}.
\newblock {\em J. Exp. Sci. Environ. Epidemiol.}, 21(1):42--48, 2011.

\bibitem{aqe}
{Tangient LLC}.
\newblock {Air Quality Egg project, airqualityegg.com }, 2013 Date of access:
  01/02/2014.

\bibitem{dynamics2013}
Francesca Tria, Vittorio Loreto, Vito Domenico~Pietro Servedio, and Steven~H
  Strogatz.
\newblock The dynamics of correlated novelties.
\newblock {\em Nat. Sci. Rep.}, 4, 2014.

\bibitem{VandenBossche2015}
Joris {Van den Bossche}, Jan Peters, Jan Verwaeren, Dick Botteldooren, Jan
  Theunis, and Bernard {De Baets}.
\newblock {Mobile monitoring for mapping spatial variation in urban air
  quality: development and validation of a methodology based on an extensive
  dataset}.
\newblock {\em Atmos. Environ.}, January 2015.

\end{thebibliography}



\section*{Supplementary material}

# Participatory patterns in an international air quality monitoring initiative
# Supplementary material

Alina Sîrbu, Martin Becker, Saverio Caminiti, Bernard De Baets, Bart Elen,
Louise Francis, Pietro Gravino, Andreas Hotho, Stefano Ingarra, Vittorio Loreto,
Andrea Molino, Juergen Mueller, Jan Peters, Ferdinando Ricchiuti,
Fabio Saracino, Vito D. P. Servedio, Gerd Stumme,
Jan Theunis, Francesca Tria, Joris Van Den Bossche

## 1 The EveryAware sensor box

All the air quality sensors (Table 1) were subjected to laboratory and outdoor tests for further characterization. In laboratory tests, sensors were exposed to synthetic gas mixtures (CO and $NO_2$) at a constant temperature of 25°C and a relative humidity of 50%. The sensors included in this analysis were the Alphasense CO-BF sensor, the e2v MiCS-5521 CO sensor and the e2v MiCS-2710 $NO_2$ sensor. The experiments were run for approximately 3 hours. The CO sensors were exposed to a series of CO gas concentrations of 9.18, 6.89, 4.61, 2.3 and 1.15 ppm. Between the different concentrations, a zero measurement was made. Each step of the measurement series lasted for approximately 20 minutes. The $NO_2$ measurements were made with concentrations of 85, 44, 24 and 0 ppb. The response times of the sensors (T90), defined as the time required for the sensor to reach 90% of its maximal value in response to a step change from zero to a certain concentration value, was monitored at a 30 second resolution. Average T90 response times were 150 seconds, 180 seconds and 270 seconds for the Alphasense CO-BF sensor, the e2v MiCS-5521 CO sensor and the e2v MiCS-2710 $NO_2$ sensor, respectively. The linearity of the sensors was high for the Alphasense CO-BF sensor in the 0 - 10 ppm CO range ($R^2 > 0.99$), and for the e2v MiCS-2710 $NO_2$ sensor for $NO_2$ concentrations between 0 and 90 ppb ($R^2=0.98$). The e2v MiCS-5521 CO sensor showed a non-linear relationship in the 0-10 ppm CO concentration range.

Because the controled laboratory setting is very different from outdoor conditions, the main tests were performed outdoor. The outdoor performance tests were carried out at a station from the Flemish air quality monitoring network. The station (Borgerhout, 42R801, see www.ircel.be) is situated at a traffic location along a double lane main street with an average daily traffic volume of 43,381 vehicles (42,961 cars and 420 heavy duty vehicles, data from the Traffic Centre Flanders). We used 4 sensor boxes for the outdoor performance tests from October 2012 until April 2013. By placing the sensor boxes at an official monitoring station we gained the advantage of having reference data for several pollutants (CO, NO, $NO_2$, $O_3$ and Black Carbon (BC)) albeit at a coarser temporal resolution of 30 minutes. The average gas (CO, NO, $NO_2$ and $O_3$) concentration and BC concentration during the outdoor tests are given in Table 2. A cross-correlation analysis was performed to compare the 30 minute averaged sensor data with the reference data for several pollutants (Table 3). Correlation between the reference data is given in Table 4 for comparison.

Low to moderate correlations were found between the CO sensor measurements and the

Table 1: Overview of the sensors of the sensor box.

| Sensor | Measured parameter | Dynamic range | Cost |
|---|---|---|---|
| Alphasense CO-BF | CO | | 180 Euro |
| e2v MiCS-5521 | CO | 1 - 1000 ppm | 3.4 Euro |
| e2v MiCS-5525 | CO | 1 - 1000 ppm | 5 Euro |
| Figaro TGS 2201 (dual) | CO | 10 - 1000 ppm | 15 Euro |
| Figaro TGS 2201 (dual) | NOx | 0.1 - 10 ppm | 15 Euro |
| e2v MiCS-2710 | $NO_2$ | 0.05 - 5 ppm | 3.7 Euro |
| e2v MiCS-2610 | $O_3$ | 10 - 1000 ppb | 3.7 Euro |
| Applied Sensors AS-MLV | VOC | not available | 5 Euro |
| Sensirion SHT21 | temp | -40 - 125 ° | 15 Euro |
| Sensirion SHT21 | rel. humidity | 0 - 100 % | 15 Euro |

Table 2: Average concentration and standard deviation of CO, NO, $NO_2$, $O_3$ and BC by the reference monitors during the outdoor testing period.

| | mean | stdev. |
|---|---|---|
| CO | 0.32 ppm | 0.13 ppm |
| NO | 32.27 ppb | 35.34 ppb |
| NO2 | 26.56 ppb | 11.50 ppb |
| O3 | 9.87 ppb | 10.00 ppb |
| BC | 4.03 $\mu g/m^3$ | 2.73 $\mu g/m^3$ |

Table 3: Cross-correlation between sensor measurements and reference gas measurements or meteorological data measured inside the sensor box. Averages of 4 sensor boxes are shown together with the standard deviations between brackets.

| | Reference monitors | | | | | sensor box | |
|---|---|---|---|---|---|---|---|
| Sensors | CO* | NO* | $NO_2$ * | $O_3$ ** | BC* | temp (°C)* | % RH* |
| Alphasense CO-BF | 0.52 (0.16) | 0.41 (0.11) | 0.34 (0.11) | -0.32 (0.14) | 0.35 (0.13) | -0.81 (0.11) | 0.00 (0.16) |
| e2v MiCS-5521 CO | 0.31 (0.04) | 0.32 (0.04) | 0.34 (0.04) | -0.09 (0.11) | 0.41 (0.02) | 0.89 (0.06) | -0.214 (0.06) |
| e2v MiCS-5525 CO | 0.60 (0.02) | 0.51 (0.05) | 0.56 (0.05) | -0.71 (0.05) | 0.55 (0.06) | 0.50 (0.06) | 0.25 (0.03) |
| Figaro TGS 2201 CO | 0.25 (0.02) | 0.32 (0.01) | 0.17 (0.00) | -0.48 (0.01) | 0.38 (0.01) | 0.45 (0.03) | 0.46 (0.07) |
| Figaro TGS 2201 NOx | -0.78 (0.01) | -0.40 (0.06) | -0.24 (0.05) | 0.47 (0.05) | -0.47 (0.06) | -0.40 (0.04) | -0.21 (0.03) |
| e2v MiCS-2710 $NO_2$ | -0.58 (0.02) | -0.40 (0.06) | -0.31 (0.08) | 0.64 (0.07) | -0.49 (0.06) | -0.40 (0.07) | -0.27 (0.02) |
| e2v MiCS-2610 $O_3$ | -0.67 (0.06) | -0.56 (0.02) | -0.55 (0.05) | 0.83 (0.07) | -0.62 (0.03) | -0.18 (0.15) | -0.12 (0.19) |
| Applied Sensors AS-MLV VOC | 0.63 (0.02) | 0.43 (0.17) | 0.53 (0.15) | -0.44 (0.26) | 0.45 (0.19) | 0.23 (0.22) | 0.14 (0.10) |

\* streetside
\*\* backyard (30 m from street)

Table 4: Cross-correlation between reference gas measurements and meteorological data measured inside the sensor box (average of 4 sensor boxes).

| | Reference monitors | | | | | sensor box | |
|---|---|---|---|---|---|---|---|
| | CO | NO | $NO_2$ | $O_3$ | BC | temp (°C) | % RH |
| CO | 1.00 | 0.77 | 0.62 | -0.55 | 0.83 | -0.07 | -0.09 |
| NO | 0.77 | 1.00 | 0.76 | -0.51 | 0.89 | 0.13 | -0.06 |
| $NO_2$ | 0.62 | 0.76 | 1.00 | -0.53 | 0.81 | 0.11 | -0.24 |
| $O_3$ | -0.55 | -0.51 | -0.53 | 1.00 | -0.54 | -0.12 | -0.22 |
| BC | 0.83 | 0.89 | 0.81 | -0.54 | 1.00 | 0.23 | -0.08 |

2

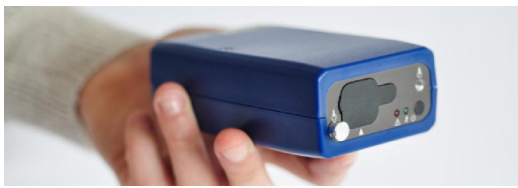

Figure 1: Micro-aethalometer: device used as a reference for calibration.

CO reference data. The Alphasense CO-BF and e2v MiCS-5525 CO sensors had the highest correlations (0.52 and 0.60, respectively) of the four CO sensors, although the Alphasense CO-BF sensor showed significant variability between sensors. The non-CO sensors showed higher correlations with the reference CO measurements. By sharing the same sources in the urban environment, it is logical that monitoring signals of different pollutants show high correlations (e.g. Table 4). The high correlations between the non-CO sensors and the reference CO data can therefore be explained. The Figaro TGS 2201 NOx and e2v MiCS-2710 $NO_2$ sensors showed a moderate correlation with NO and $NO_2$. The negative sign is due to the electronics and can be discarded in this analysis. Correlations of the NOx sensors are higher with CO and $O_3$, although these values stayed within the correlation range that was observed for the reference measurements. It is not proven by this experiment that the high correlation with CO and $O_3$ is due to selectivity problems of the sensor. The e2v MiCS-2610 $O_3$ sensor showed a high correlation with the reference $O_3$ measurements. Also the variability between the sensor boxes was limited. Correlations with other gases are negative, which is in line with the physical reality (see reference measurements in Table 4). The Applied Sensors AS-MLV VOC sensor shows the highest correlations with CO and $NO_2$, but reference VOC measurements are lacking. The correlation with BC ranges between 0.35 and -0.62. These correlations are in the same moderate range as the correlations that are found between the sensors and the reference measurements of the respective gases.

Important conclusions of these laboratory and field experiments with respect to further developments and applications of the sensor box are: (i) response times of the sensors are in the minute range rather than in the second range; (ii) correlations between sensor and reference measurements are low to moderate for most of the sensors, for the e2v MiCS-2610 $O_3$ sensor the correlation is high; (iii) moderate correlations between the sensor measurements and the BC measurements are observed.

## 2 Sensor box calibration

Following the analysis of sensor abilities presented in the previous section, we have proceeded with calibration of the sensor boxes. Issues identified by our initial analysis included sensor sensitivity to temperature and humidity, sensor drift in time and sensitivity to other gasses. Hence, one needs to calibrate devices against a reference in order to control for this issues and obtain a measurement meaningful for the user.

Supervised learning was employed to model an unknown concentration of a target pollutant from sensor array measurements (based on low-cost gas sensors, temperature and relative humidity sensors). The supervised learning model is parameterised by using a training dataset consisting of sensorbox measurements and simultaneous target pollutant concentration measurements. The target pollutant selected in this study was black carbon (BC). The selection of BC as a target was motivated by following three reasons: (1) BC is a relevant pollutant in urban

Table 5: Performance of the ANN, SVM and RF techniques to model BC from sensor box data on independent mobile data.

|     | $R^2$ | rmse |
| --- | --- | --- |
| ANN | 0.26 | 2.10 |
| SVM | 0.23 | 1.70 |
| RF  | -0.13 | 1.72 |

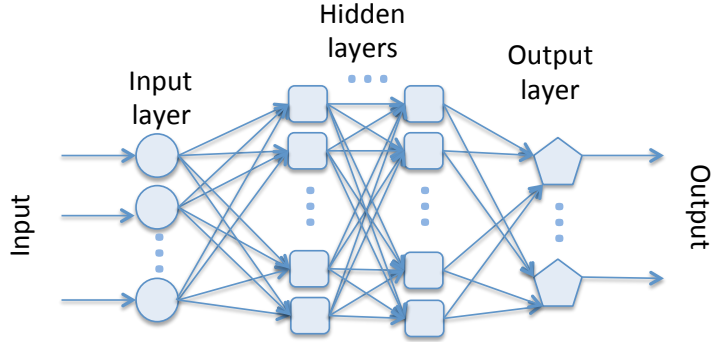

Figure 2: General structure of artificial neural networks.

environment by its adverse health effects [1], (2) BC is correlated with the gases that are measured by the sensor box, as seen in the previous sections, and (3) the availability of portable BC measurement devices (micro-aethalometers, AethLabs, Figure 1) which makes it possible to collect mobile BC data.

## 2.1 Calibration model

The first calibration datasets, consisting of sensor box data measured at the same time with micro-aethalometer data, were obtained in Antwerp near an air quality monitoring station at a traffic site and in Turin in spring 2013 from a two-week long monitoring with sensor boxes and micro-aethalometers positioned near a busy road. These datasets were used to compare the performance of different supervised learning techniques. We have explored four different possible models to use for mapping of sensor output to the reference measurements. These were Random Forests, Support Vector Machines (SVMs), a custom air quality index and Artificial Neural Networks (ANNs). After comparing these (Table 5), SVMs and ANNs obtained a similar behaviour, better than the other two options, but training of ANNs appeared to be faster, so we decided to adopt them for our model.

ANNs[2] are regression models that mimic the behaviour or neuronal networks. They consist of interconnected computing units (neurons) that can have several inputs and an output. In each unit, two operations happen: compute a weighted sum of the inputs and apply a sigmoid (activation) function to obtain the output. The network can have several layers feeding into one another: one input and one output layer plus a number of hidden layers (see Figure 2).

In order to train a network, one needs to select a topology and find the values of the input weights for each neuron. To select the topology we have performed an empirical analysis that led to the usage of a network with one hidden layer of 10 neurons. The final topology is displayed in Figure 3. After this, we used backpropagation, a standard algorithm for ANN training [2], to

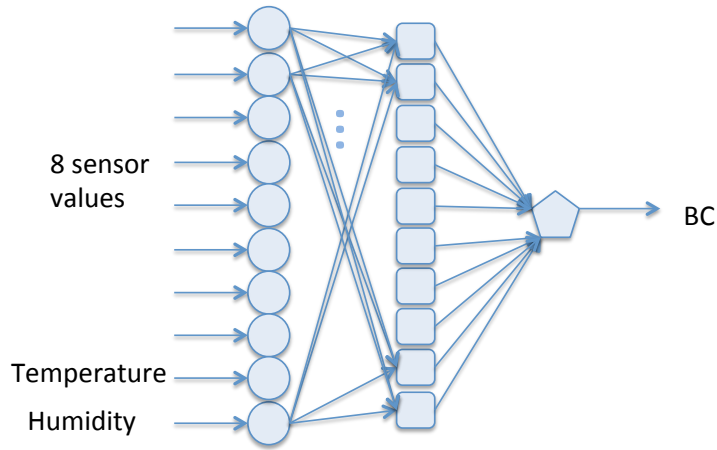

Figure 3: ANN topology for our calibration problem.

obtain the network weights.

The ANN model was implemented both in the AirProbe mobile application as well on the server. This was done to give the user real time feedback from the sensor box on the one hand, but on the other hand allow the mobile device to leave the computationally expensive calculation of the black carbon value to the server in case the user is not using the viewing of black carbon values in real-time.

This approach was taken due to the fact that the sensor box has two working modes, online and offline. Computing model output for all offline records would have been too computationally expensive for an average smartphone, while server side this was not an issue.

## 2.2 Preprocessing

Although initially the possibility of building one calibration model for each box was intended, this would not have scaled very well, so we explored the possibility of building one model for all sensor boxes. From the first calibration data sets we observed that sensor boxes behaved similarly when exposed simultaneously. Although the absolute sensor values differ between boxes, the fluctuations in time are similar for the same sensors. This means that sensor box rescaling (i.e. normalisation of the sensor signals) could be used to scale the different sensor boxes within the same range, and parameterise a model on the standardised data. This would mean obtaining a unique model for all sensor boxes, instead of individual ones for each box. For APIC we decided to use one calibration model for each city to account for possible differences in sensor response between locations.

A different issue was data variability, both in BC values and sensor response. The BC values were first processed by a noise reduction algorithm [3] to lower the high-frequency instrument noise that is observed when measuring at high frequency. To remove sensor box fluctuations, the sensor data was smoothed by computing averages over a moving time window of one minute. The resulting BC values were averaged over a 5 minute window. This value was deemed suitable by comparing outputs from two aethalometers, which become highly correlated at this resolution. So the BC value obtained from the model represents an average over the last five minutes of exposure.

## 2.3 Model performance

For each location of the challenge, models were trained on experimental data. Three types of data were used, to account for three possible use cases. These included stationary data where all sensor boxes were collocated, mobile measurements performed with one or two boxes at a time, and indoor data. Training and testing datasets were obtained by combining all these data types, and four models were obtained, one for each location.

Figures 4, 5, 6 and 7 display the best model obtained in each city, in terms of performance on training and test data, as well as cumulative exposure on the test data. The Turin model (Figure 7) performs best, and this is due to the increased amount of data, especially mobile, available for this location, since preliminary calibration tests were performed there. This demonstrates that the collection of large amounts of mobile data is crucial for boosting modelling abilities. The Antwerp dataset also contained larger amounts of mobile data, compared to the other two, so that a good performance was obtained as well (Figure 4). Although datasets were more restricted for Kassel and London, indications were that models obtained were displaying good performance (see Figures 5 and 6).

In general, calibration was successful at identifying general trends in the pollution levels. However, sharp and short peaks are not handled well by the model, and this is due to the lower sensitivity of the low-cost sensors and their delayed response. However, the performance obtained was enough for the purposes of our project, i.e. participatory mapping of pollution with multiple devices, for enhancing environmental awareness.

# 3 The AirProbe application

The AirProbe application is freely available for the Android platform and can be installed from Google PlayStore. The main objective of the application is to acquire the data from the sensor box and to upload it to the EveryAware server. The application also allows the user to view/annotate data and can operate in three different modes: Live Track, Synchronization and Browsing. Without this application, the sensor box data cannot be accessed nor uploaded to the servers.

In order to associate the data uploaded with a specific user, the application must be activated. This process links the application to an existing EveryAware account (which can be created inside the application itself or on the project web site).

## 3.1 Live Track mode

This is the standard way to use AirProbe to collect air quality measurements. In this mode, the application will search for Bluetooth devices nearby and present the user with a list of found devices. EveryAware sensor boxes can be easily identified by their MAC suffix. Once the user has selected the sensor box, AirProbe starts displaying real time data collected by the sensor box, using the Bluetooth connection. In Live Track mode, the interface is composed of three different views accessible from their corresponding tabs (Figure 8):

**Map** , where users can follow their own live track. The track is represented with different colours, depending on real-time black carbon levels. The user can also add annotations and share them on social networks (Facebook/Twitter), using the buttons at the top right corner. The track length to be shown on the map can be of 5, 15, 60 minutes. Live tracking of the current position can be switched on/off, through the top left buttons. The bar at the top represents the black carbon value using a coloured scale (from a blue/low value to a brown/high value).

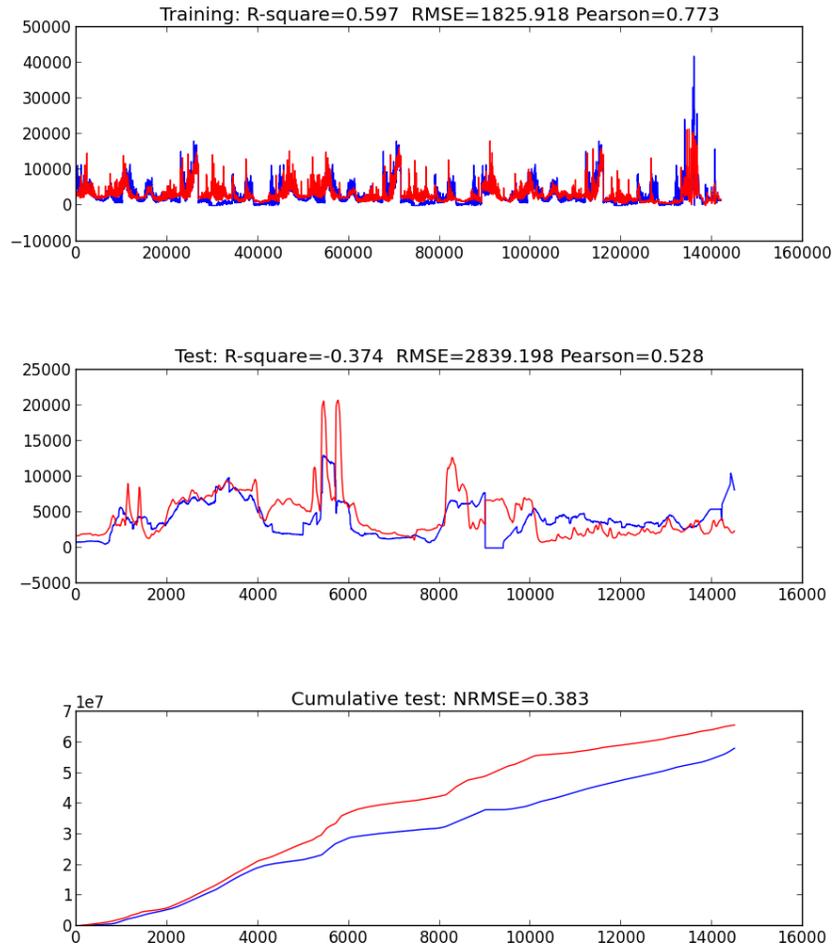

Figure 4: Model performance in Antwerp. The red line represents the model, while the blue line represents the real data from the reference device.

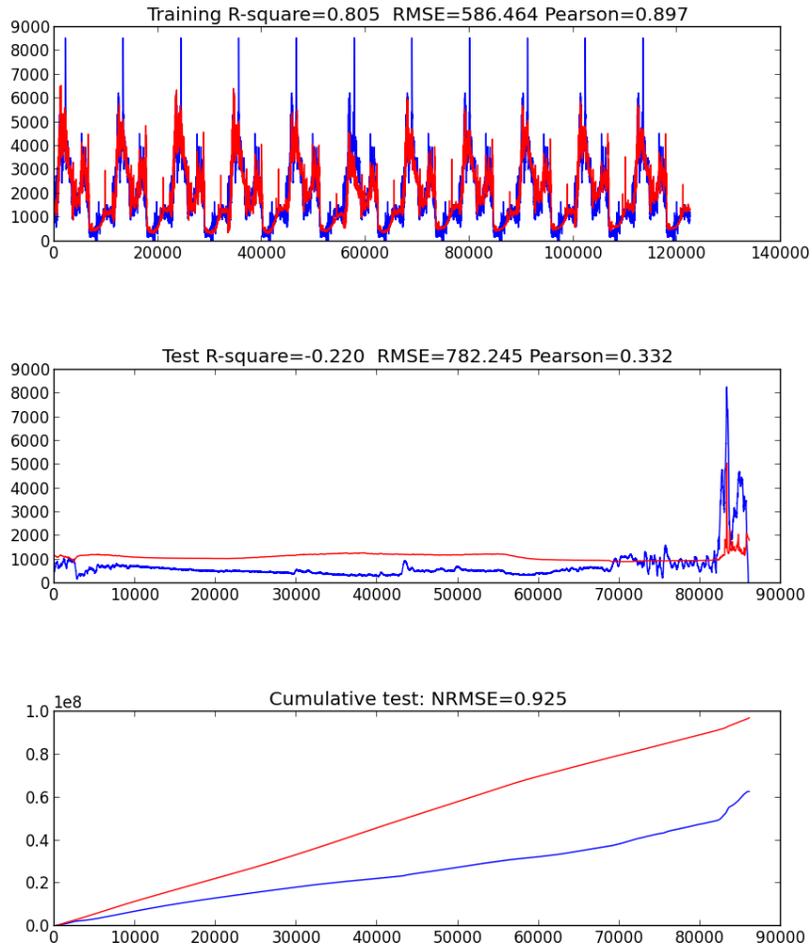

Figure 5: Model performance in Kassel. The red line represents the model, while the blue line represents the real data from the reference device.

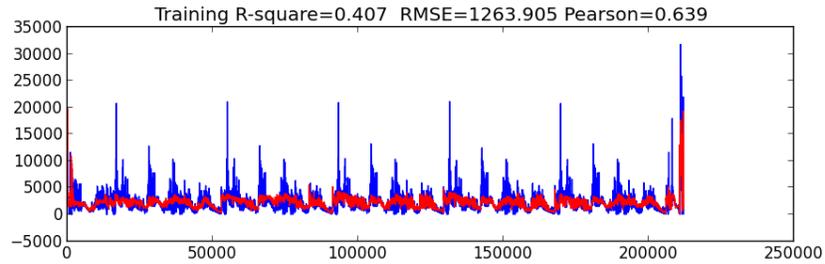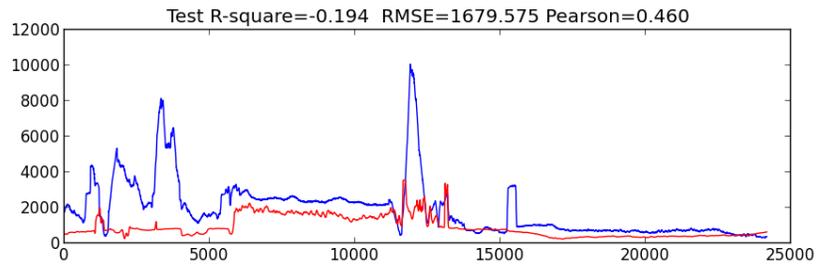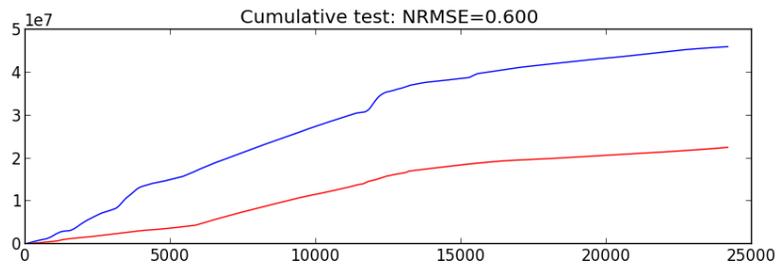

Figure 6: Model performance in London. The red line represents the model, while the blue line represents the real data from the reference device.

9

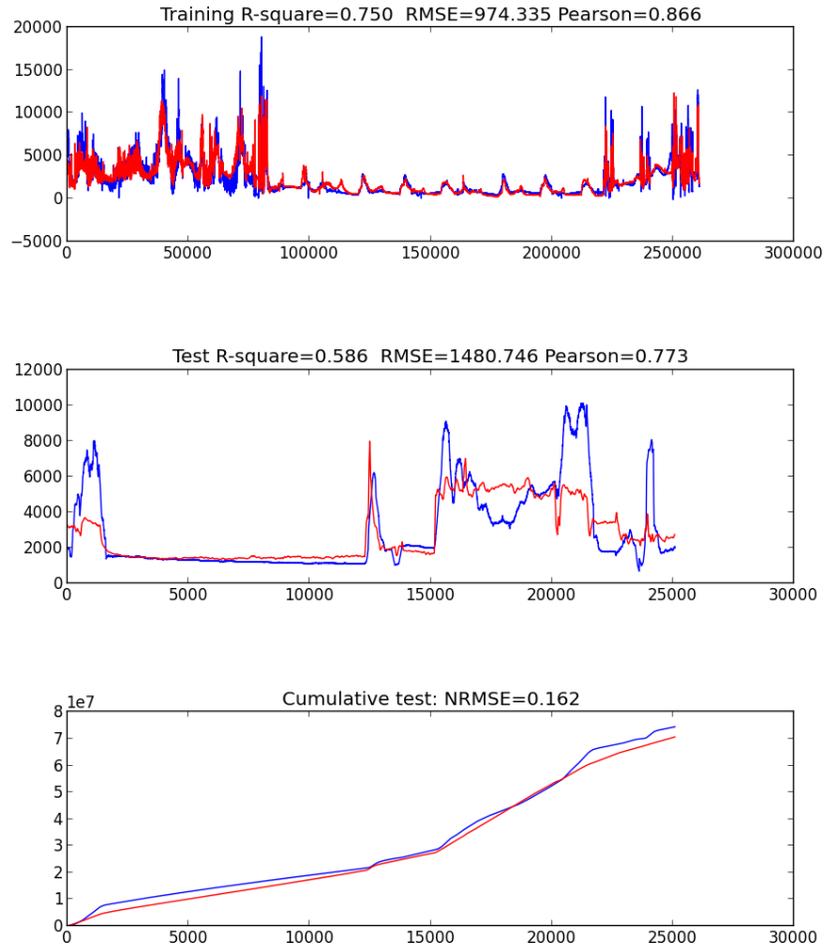

Figure 7: Model performance in Turin. The red line represents the model, while the blue line represents the real data from the reference device.

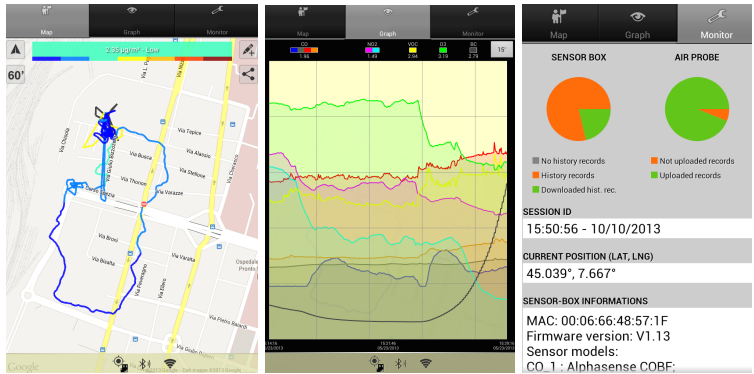

Figure 8: AirProbe screenshots: Live mode. AirProbe uses the Google Maps API to display maps (©2014 Google [4]).

**Graph** , where the user can see black carbon evolution and the raw data from pollutant sensors, in a variable time interval ranging from 1 to 30 minutes. The user can query the value registered by each sensor by tapping on the series. The graph is updated every two seconds.

**Monitor** , where users can access statistics about collected data, connection information, the status of the sensor box and the installed sensors.

## 3.2 Synchronization mode

In this working mode, AirProbe downloads data from the sensor box and uploads them to the EveryAware server (Figure 9). The sensor box in this case is used as a pure data logger, allowing the user to send data only in suitable conditions (e.g. where battery lifetime and/or connection billing are not a problem).

## 3.3 Browsing mode

This working mode does not require an active Bluetooth connection to a sensor box. It is composed by three views, accessible from their corresponding tabs:

**Map** , where the user can see the black carbon levels around his current position (Figure 9), by pressing the "Get nearby BC levels" button. If a track from "MyTracks" tab is selected, this is displayed on the map. The black carbon levels and selected track can be shown together.

**Graph** , where the raw pollutant and black carbon evolution, calculated for a selected track, are shown. Only live recorded tracks have black carbon data.

**My Track** , where the list of tracks available on the mobile device is shown. Older tracks are automatically deleted only once they have been uploaded to the server and a configurable time interval since their creation has passed.

11

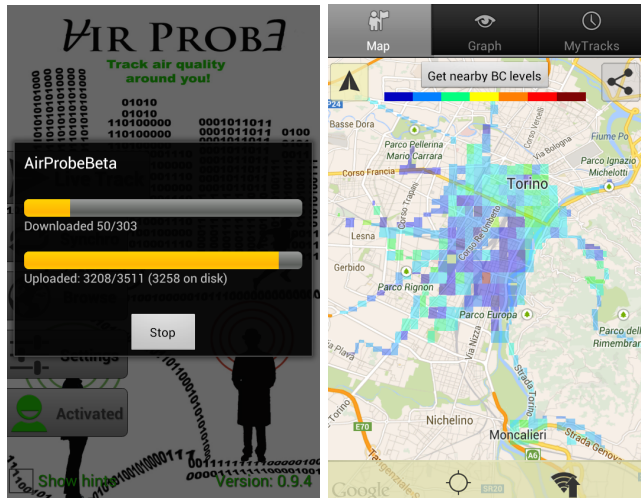

Figure 9: AirProbe screenshots: Synchronization and Black Carbon map. AirProbe uses the Google Maps API to display maps (©2014 Google [4]).

## 4 The web platform

The personalized air quality data is collected using the AirProbe module of the EveryAware platform which is a social information technology system used by people to communicate and share information. A characteristic of social information technologies is that they often involve very large amounts of data. In fact, the collection, storage, and analysis of different kinds of data within these systems is a crucial point and also an asset, e.g., for companies like Facebook[1]. As a consequence, in order to pave the way towards analyzing and even triggering behavioral shifts within large citizen populations, methods and techniques of acquiring and handling such data efficiently play a central role. The design of web-based infrastructures for this purpose has a great influence both on data quantity and quality, and hence also on the additional value which can be generated by analyzing the resulting datasets. Typical goals during the design process are:

- *Performance:* All infrastructure modules must be carefully tuned for high-performance requirements of processing large amounts of data in a parallel fashion because the involvement of large numbers of users requires responsive interfaces and efficient server backends

- *Management:* The setup and technical realization of experiments and studies among citizens often implies strong efforts on the side of scientists and experimenters. As a consequence, it is desirable to provide reusable and configurable experimentation platforms which can easily be managed.

- *Correctness:* A large-scale collection of data can hardly be expected to provide only correct and consistent results. However, the reduction of noise from the very beginning (i.e., the concrete measurements) is desirable in order to provide a better basis for later analysis.

---

[1] http://facebook.com

Broadly speaking, the relevant data in the context of the EveryAware project can be divided into two classes, namely

1. *objective data*, which stems mainly from sensors and captures information like sound intensity or air quality measurements (as analyzed in this work), and

2. *subjective data*, which comprises context information about the collected data including reactions of humans faced with particular environmental conditions. This comes from annotations that users attach to their measurements and is different from the subjective data we collected through the web game.

The EveryAware platform has been explicitly designed to support subjective impressions in conjunction with sensor data acquisition by introducing an extendable data concept. A central server efficiently collects, analyzes and visualizes data sent from the arbitrary sources. The platform offers a highly flexible way to store and exchange data for Internet of Things applications. A wide variety of meta, location, and content information which can be attached to any data point, a flexible *data processor component* as well as an efficient storage structure are the keys to this task. These mechanisms provide the unique ability to enrich data with contextual information explicitly including subjective impressions. Different collection concepts like sessions to represent time-interval-based entities and feeds to organize data points in a continuous way allow to further introduce semantic relations. This enables the web interface to provide different semantically enriched views on the data, aggregating data globally as well as on a personal level. Allowing users to access this information is a crucial part of the system since it closes the loop from data collection to analysis to pushing information back to individual users and communities which in turn triggers new collection activities. For more information about the EveryAware platforms and its components we refer to Becker et al. [5].

## 4.1 Statistics and visualizations

In the case of AirProbe the visualized information is represented by several views of the data including a map with different information layers as well as several global and personal statistics. The OpenStreetMap-based[2] map view visualizes the collected data on a map which allows for an easy access to the data as well as for obtaining first insights. It provides a quantitative view by aggregating samples using clusters, grids, as well as a heatmap view in order to emphasize the covered area on a global and on a personal level (see Figure 10).

Further statistics calculated by the AirProbe application include summaries like latest overall measurement activity or air quality averages. Also, personal user profiles are available which list measurement sessions giving short summaries regarding those sessions and enabling the user to view and replay them. A personal sessions overview can be seen in Figure 11(a). One view for exploring personal sessions can be seen in Figure 11(b).

## 4.2 APIC rankings

Additionally, the web interface provided feedback for the users participating in the APIC game by measuring air quality using sensorboxes. The case study was held in order to gather large amounts of air quality samples and behavioral shift patterns using the sensorboxes in the four cities Antwerp, Kassel, London, and Turin.

In order to keep the motivation and competitiveness as high as possible for the teams playing, we implemented a ranking mechanism balancing repetitive sampling and coverage. The map was

---

[2]http://openstreetmap.org/

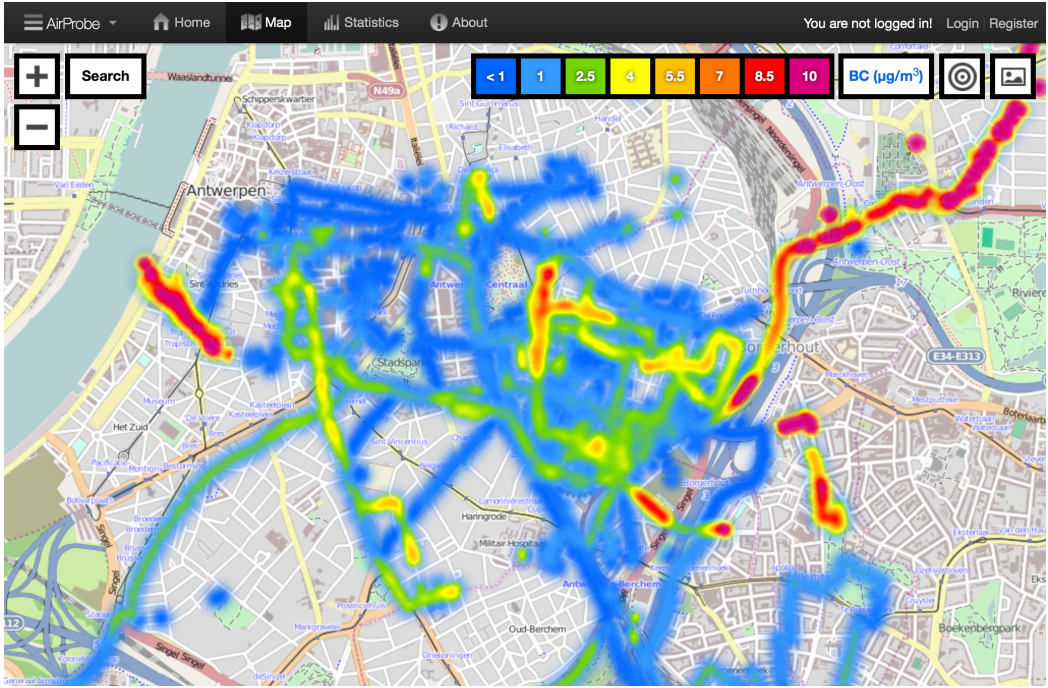

Figure 10: A screenshot of a heatmap on the *map page* of AirProbe. The website map and heatmap were generated using in-house developed tools and OpenStreetMap data (©*OpenStreetMap contributors* for map data, used and redistributed under the CC-BY-SA licence[6]) .

divided into 10 by 10 meter grids. One point was given to a team when sampling within one such grid cell. When a team received a point in a particular cell, the player did not receive a point from this grid cell for half an hour. The results for each city as well as for each team have been visualized and updated in regular intervals on the AirProbe website as can be seen in Figure 12. Figure 12(a) shows the ranking of each city visualizing the coverage and providing several statistics. Figure 12(b) shows a detailed view of the point-coverage of the city.

## 5 The APIC web game

In order to gather subjective opinions about air pollution in the four cities we decided to follow the *game with a purpose* [7] approach and accomplish the task using a web game. We started designing the game taking inspiration from the specific kind of data we wanted. Our aim was not only to get a map of perceived air pollution but also to study how the perception is affected by objective data. Specifically, we needed to monitor volunteer opinion before, during and after exposure to objective air quality data, obtained by the sensing device. This meant keeping the players engaged in the game for the longest time possible, in order to monitor the opinion shift of each player. Beside this, opinions about air pollution had to be geo-localised so the game had to take place on the maps of the four cities. In particular, for each city we defined a mapping area of approximately 3 km². The mapping areas are represented in Figure 13.

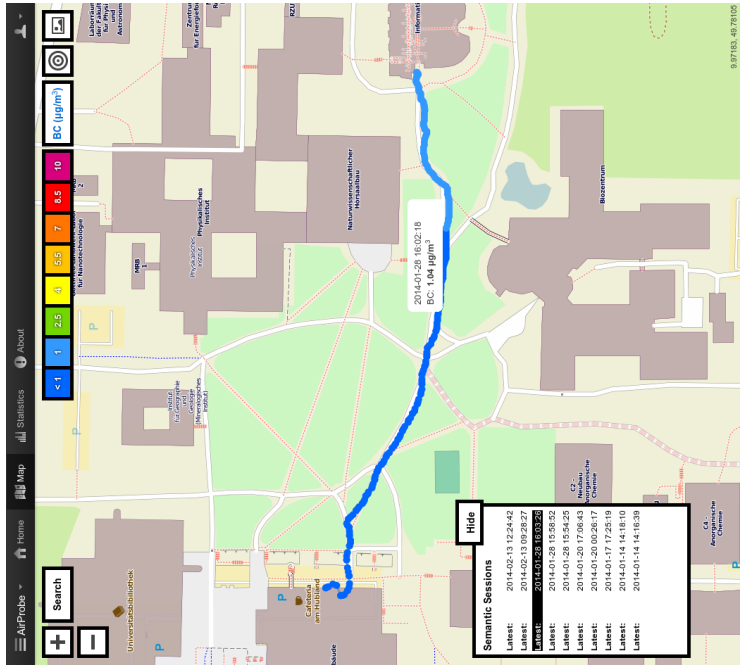
(a) This AirProbe view shows a user's personal sessions.

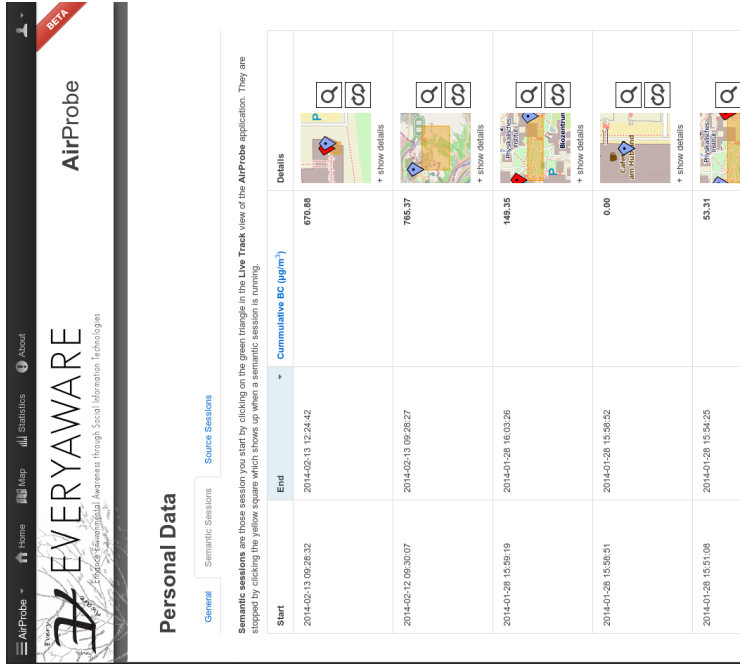
(b) This AirProbe view shows a view for exploring individual user sessions.

Figure 11: AirProbe personal measurement sessions visualizations. The website map and track visualisation were generated using in-house developed tools and OpenStreetMap data (©*OpenStreetMap contributors* for map data, used and redistributed under the CC-BY-SA licence[6]) .

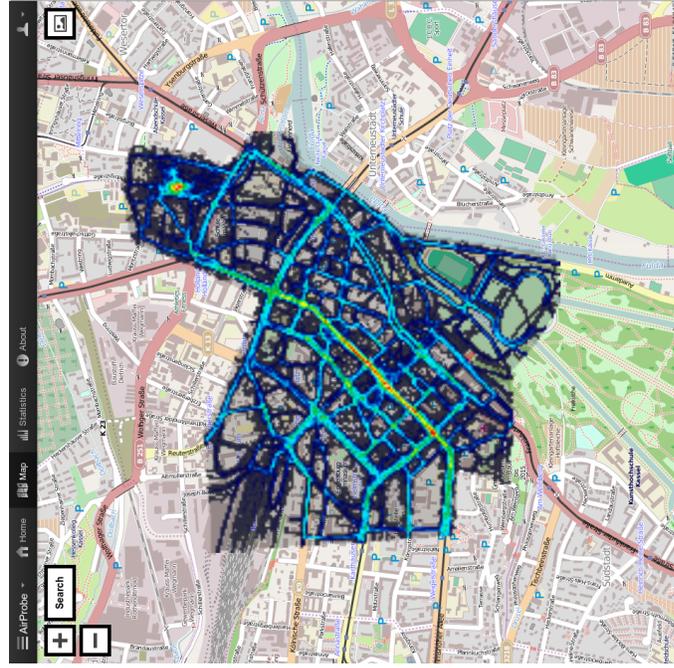

(a) APIC city ranking.

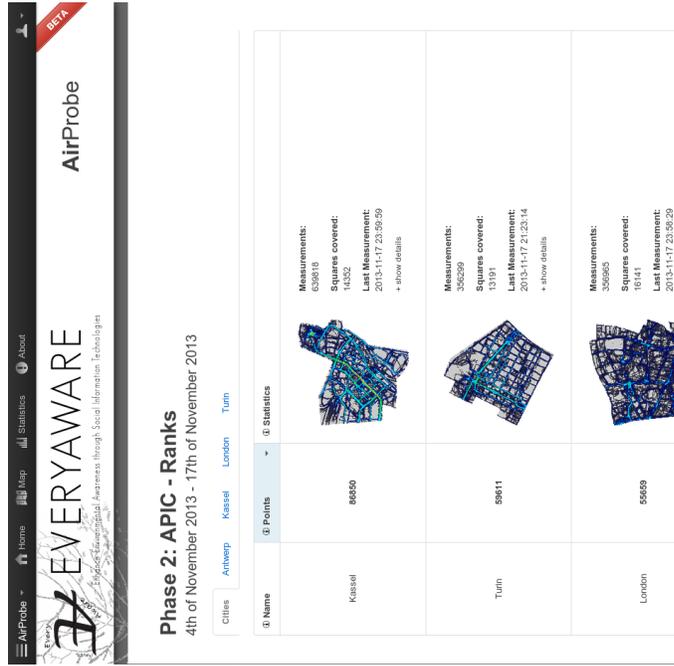

(b) APIC point coverage for Kassel.

Figure 12: APIC ranking visualizations. The website map and heatmaps were generated using in-house developed tools and OpenStreetMap data (©*OpenStreetMap contributors* for map data, used and redistributed under the CC-BY-SA licence[6]) .

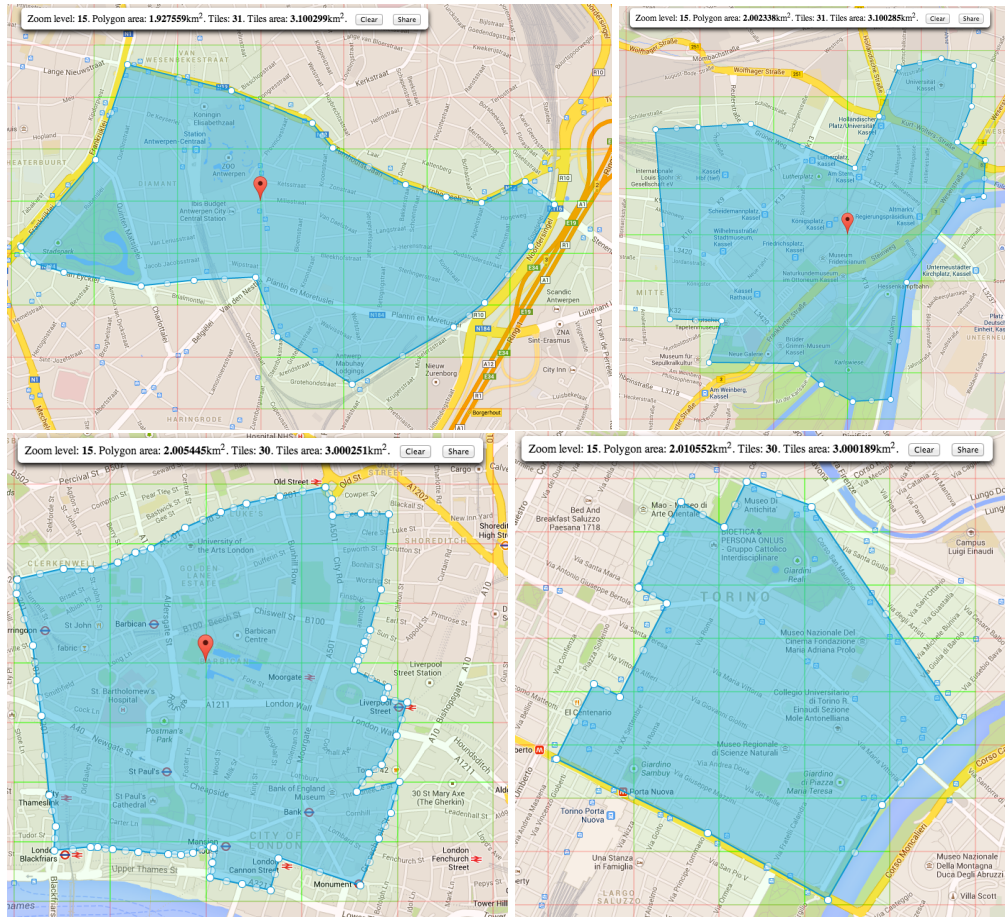

Figure 13: In green the game areas and in blue the measurement areas for the four cities. The grid represents the tiles division for the web game. From the top left to the bottom right: Antwerp, Kassel, London and Turin. The images were generated using the Google Maps API - polygons and screenshots (Kassel: ©2014 GeoBasis-DE/BKG (©2009) Google, other locations: ©2014 Google [4]

Considering all this, the most suitable type of game appeared to be a management simulation, like the famous FarmVille or Harvest Moon. In this type of game the user has the task of managing a given territory. By improving their management performances, the users increase their income in the game. Thus they may access a wider set of features, for example they can expand their territory or buy more objects, all in order to further improve their income. The periodic rhythm of this cycle is marked (in FarmVille-like games) by the time the income is claimable by the player: in order to generate a revenue, an action is required at a given time, spanning from a few seconds to several hours. This mechanism is an incentive to return to the game, in order to gather the results of one's effort.

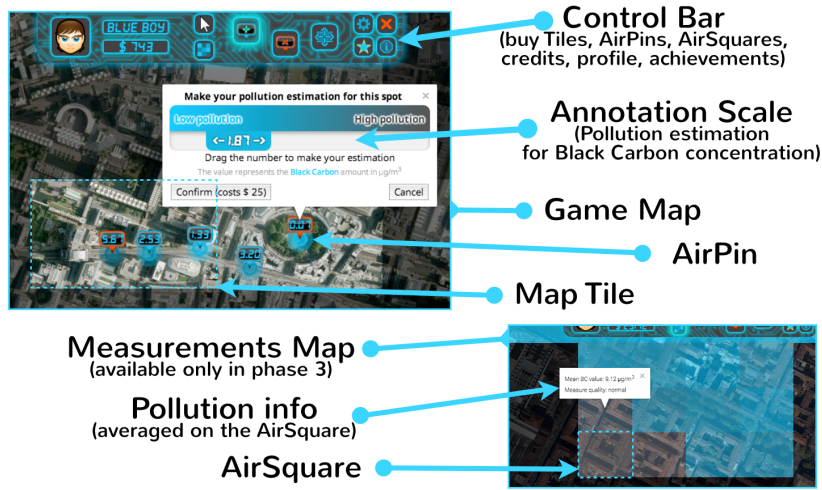

Figure 14: Screenshots of the game interface, with indication of the main entities and tools. The game uses the Google Maps API to display maps (©2014 Google [4]).

The AirProbe web game is a simplified map management game. In Figure 14 the interface of the game is depicted. Players are called to fulfil their role of Air Guardians by annotating the map with AirPins: geo-localised flags tagged with an estimated or perceived pollution level (Black Carbon concentration in $\mu g/m^3$, on a scale from 0 to 10). The game area of each city is divided into tiles as indicated in Figure 13.

At the beginning of the game, users are asked to create a profile (by choosing an avatar and a name) and to choose a city and a team. Teams were linked to Air Ambassadors, and were an important part of the competition. Then the volunteer starts from a given Tile of the map of the chosen city. The user can interact by placing (or editing or removing) AirPins or by expanding their territory by buying more Tiles. Each day the AirPins placed generate a revenue based on the precision of the annotation (more details in the following). In order to collect the revenue generated every day by each AirPin, the user has to access the game daily, otherwise the revenue will be wasted. The revenue collected will be added to the user balance, and can be used to buy more AirPins and more Tiles. In order to improve motivation and fidelity, there is a bonus for days-in-a-row accesses and a large set of other achievements. These achievements consist of prizes at given milestones in the game story: a certain number of AirPins or Tiles, precision in the annotation, and so on.

In phase 3 of the case study we made available information about objective measurements

gathered with the sensor box during phase 2. We avoided to give punctual information about measurements, otherwise it was likely that users would simply copy the values. So we decided to release aggregated information by introducing a new map partition named AirSquares. Each Tile contains 12 AirSquares, that can be purchased just like AirPins or Tiles. Once users bought an AirSquare, they can see the average pollution value in that area, so the task changes into estimating fluctuations.

## 5.1 Revenue and feedback

Our case study was divided in three phases. Beside the AirSquare introduction in phase 3, the only change between phases was in the revenue calculation algorithm. We generically said that revenue was related to precision of the annotation. Let us now define the meaning of 'precision' in our context. In phase 1 there were no objective data for comparison, thus we adopted the strategy of matching the AirPins with the estimations of other users within a certain range (30 meters).

The general algorithm of revenue calculation for a certain AirPin $f_0$ with coordinates $(lon_0, lat_0)$ and value $bc_0$ was chosen in order to fulfill these conditions:

**No data** Even if we solved the problem of the lack of data by comparing a user annotation with other user annotations, at the beginning of phase 1 those were missing as well. So, in case of absence of other AirPins within the range, the only choice was to trust the user and give him an average revenue for the AirPin.

**Distance** In case other AirPins do exist within the range, their distance from the location of $f_0$ had to be taken into account.

**Reliability of the match** If an AirPin value matched those of a large number of other AirPins, the revenue had to be large. So the maximum possible revenue is determined by the number of AirPins within the range.

We decided that the most simple and reasonable choice to give revenue for an AirPin $f_0$ was based on a comparison of the Black Carbon value $bc_0$ associated with $f_0$ and the average of all the AirPins $F$ (including $f_0$ itself) within range of $f_0$ weighted by their distance to $f_0$ and rescaled depending on the number of AirPins in range. So, let $F = \{f_0\} \cup \{f_1, \ldots, f_n\}$ be the set of all AirPins within 30 meters from $f_0$, including $f_0$ itself and consider the tuples $(bc_i, dist_0(f_i))$ of Black Carbon estimates $bc_i$ and distances $dist_0(f_i)$ from $f_0$ for all AirPins in range $f_i \in F$. Let $bc_F$ be the weighted mean of all values $bc_i$ in $F$ using a weight $w_i$ defined as

$$w_i = 1 - \frac{dist_0(f_i)^2}{30^2} \tag{1}$$

Let $W$ be the sum of all $w_i$. Now, we computed the maximum revenue for an AirPin $f_0$ based on this sum of weights $W$. We use an inverse exponential function to adjust the maximum revenue ($r_{max}$) from 30 (when $W = 1$) to about 65 (when $W = 10$) to 75 (when $W > 20$):

$$r_{max} = 30 + 45(1 - 2^{-\frac{W-1}{3}}) \tag{2}$$

We now define the 'error' $e_0 = |bc_0 - bc_F|$ of the estimation for the AirPin $f_0$ as the absolute value of the difference between the AirPin value $bc_0$ and the weighted average $bc_F$ of $F$. Finally, we defined a critical threshold $t$ for the error. If $e_0 > t$ then the revenue will be 0, otherwise the

revenue is calculated using a formula taking the maximum revenue $r_{max}$, the error $e_0$ as well as the defnied critical threshold $t$ into account:

$$r_0 = \begin{cases} 0, & \text{if } e_0 > t \\ r_{max}\frac{1-e_0}{t}, & \text{else} \end{cases} \quad (3)$$

The users only had an aggregated view on their revenue, i.e., only the cumulative value for the whole ensemble of their AirPins was shown. The only feedback regarding single flags was a red sign for flags that were not generating any revenue.

As we said, the revenue algorithm has been different in each of the three phases:

**Phase 1** The threshold for the error was very tolerant (5 $\mu$g/m$^3$) in order to make the game easy at the beginning.

**Phase 2** The threshold was made smaller (2.5 $\mu$g/m$^3$), in order to make the game more challenging and keep users engaged.

**Phase 3** The threshold was unchanged but real measurements from sensor boxes were used instead of other players annotation to calculate the revenues.

Users were not informed about the details of the algorithm. They were just asked to try to be precise. Every day ranks were published. In order to boost motivation, we introduced a set of prizes to be given at the end of each phase and in each city. We considered two main metrics for the ranks: the total revenue of the last day of play and the number of days played in each phase (fidelity).

## 6 Recruiting activities

In order to recruit participants for the study each city adopted their own recruitment strategy alongside publicity via the APIC Facebook page, Twitter and the project website which was used across all the cities. University mailing lists were used to recruit volunteers in each location, excluding Antwerp, who alongside Kassel were the only cities to use external email mailhosts. In Antwerp, where similar air quality monitoring activities have previously been carried out, the challenge was advertised via a specific mailing list which included volunteers from earlier monitoring campaigns, traffic organisations, environmental agencies and interest groups, and communities working on sustainability issues. The advertisement included a link to a participation form that included several questions which were used to gain some ideas on the degree of interest of the participants in air quality monitoring and on the potential temporal coverage expected from monitoring activities. Kassel was the only city to release newspaper articles as part of their recruitment strategy; Turin and London gave talks during classes and a varying number of posters were distributed within university campuses in Kassel, Turin and London.

Interested individuals were asked to contact the relevant project team members and following the initial call for participation meetings were scheduled, specifically for Air Ambassadors, to explain the study in more detail and to provide guidance on using the sensor box. A summary of the number of Air Ambassadors recruited for each location is detailed in Table 6 below. The results show that using existing mailing lists, whether within a university or across other networks, was the most successful approach to securing volunteers in three of the four cities. In Turin, however, public talks proved to be the most successful.

|  | Antwerp | | Kassel | | London | | Turin | |
|---|---|---|---|---|---|---|---|---|
| Method | Responses received | Final volunteers | Responses received | Final volunteers | Responses received | Final volunteers | Responses received | Final volunteers |
| Mailing list | 32 | 19 | 7 | 5 | 48 | 30 | 8 | 2 |
| News paper | - | - | 3 | 2 | - | - | - | - |
| Talks | - | - | 2 | 1 | 1 | 1 | 11 | 8 |
| Posters | - | - | - | - | - | - | - | - |
| Other | - | - | - | - | - | 4 | - | - |

Table 6: APIC recruitment methods and resulting volunteer participation across the four cities.

## 6.1 Incentives

One of the aims of the AirProbe International Challenge (APIC) was to investigate participation patterns of volunteers for environmental monitoring studies via a web-based *game with a purpose* [7] and competition approach, combining online and offline activities. In addition, comparison of the various incentives offered across the four Case Study areas (London, Antwerp, Kassel and Turin) was explored.

The incentives offered to Air Guardians in each city were as follows: the player with the highest revenue at the end of each phase received a backpack; those ranked second to fourth t-shirts and the most active players also received a backpack. The winning metrics, as outlined in the earlier section, were calculated based on the revenue generated by the AirPins in the last day of play of each phase and fidelity based on the largest number of consecutive days played. Deviations from these incentives were made in Turin where five t-shirts and one backpack were offered for the highest revenue and three t-shirts and one backpack for fidelity. In Kassel an additional incentive was offered to Air Guardians based on the best precision (3 x €50 Amazon vouchers) and the largest most active team with at least three active members playing over a minimum of 21 days (3 x €50 Amazon vouchers). In Antwerp there were no specific incentive schemes and the only place in which prizes were mentioned was on the webpage which stated prizes were on offer for participants taking measurements and for the most active and best gamers.

In each of the four cities all Air Ambassadors were given solar panel backpacks for their contributions and variations across the four cities were as follows:

**In London** all were given T-shirts and shared £100 Amazon vouchers between each team (a total of 10 teams varying in size from two to six participants). The team who obtained the best temporal/spatial coverage won a sensor box in phase 2 and the winning team overall, defined as having the best temporal/spatial coverage and the largest number of active Air Guardians, received £400 in Amazon vouchers.

**Kassel** adopted a stricter criterion which in phase 2 offered twelve lots of €50 Amazon vouchers for those who carried out one hour of monitoring for at least seven days; twelve lots of €20 vouchers for those who carried out one hour of monitoring for at least seven days and who fell within the top 50% best ranked Ambassadors world-wide and €250, or a sensor box, for the best temporal/spatial coverage and at least 1.5 hours of monitoring completed for nine days. In phase 3 €250 Amazon vouchers were offered to the Ambassador with the best temporal/spatial

coverage and 1.5 hours of monitoring for nine days; €50 for second place and €20 for third with the best temporal/spatial coverage and at least one hour of monitoring for seven days

**Turin** offered T-shirts to all Ambassadors and two sensor boxes (one for phase 2 and one for phase 3) to those with highest coverage. In addition they gave final prizes (3 Amazon vouchers €75, €50 and €25 ) to Ambassadors with the best performing teams, using a combined criterion for evaluation (number of measurements, coverage and game activity for the Ambassador's team). These final prizes were however not advertised to the participants before the end of the challenge.

**Antwerp** did not specify any specific reward scheme.

The overall challenge winner across all cities was offered EveryAware T-shirts for their effort.

The influence of the different strategies regarding incentives are somewhat visible when analysing the data in the next section.

## 7 Data analysis

|  | Total geo-localised | Additional without location | Antwerp | Kassel | London | Turin |
|---|---|---|---|---|---|---|
| Number of measurements | 6,615,407 | 3,326,956 | 318,537 | 2,929,345 | 1,115,828 | 1,592,912 |
| Number of tags | 742 | 16 | 3 | 32 | 606 | 11 |

Table 7: Number of measurements and tags during the test case. Details for each of the four locations.

The analysis presented in this work is based on a large amount of air quality measurements collected using the EveryAware sensorbox during four weeks (phases 2 and 3 of the AirProbe International Challenge) in four European cities. Table 7 summarises the number of data points and tags collected. This shows that in Europe there were over 6 million measurements performed, with Kassel displaying the largest activity. This could be due to the fact that volunteers in Kassel were offered significant monetary rewards for their activity, unlike the other locations. The number of annotations is largest in London, which is due to the instructions they received which underlined the need for subjective annotations.

For further insight into the range of measurements obtained, we show daily (Figure 15) and hourly (Figure 16) numbers of measurements at each location. The different cities show different behaviour. In Antwerp and London, the activity decreased significantly during phase 3 of the challenge. This shows that users were mostly interested in mapping the main area of interest in the challenge. In Turin, however, activity increased during phase 3, which may indicate that volunteers were particularly interested in a different area than that chosen for phase 2, and in monitoring their own daily exposure. In Kassel, activity is more or less stable, with a slight decrease in the last week.

Daily patterns (activity per hour) show afternoon peaks for each location. For Kassel and Turin there is a significant amount of data collected during the night, showing an increased interest again in monitoring and collecting a large number of points.

Besides the general number of measurements coverage patterns are also important. The main text discusses overall coverage, both in time and space. Here we provide more details for

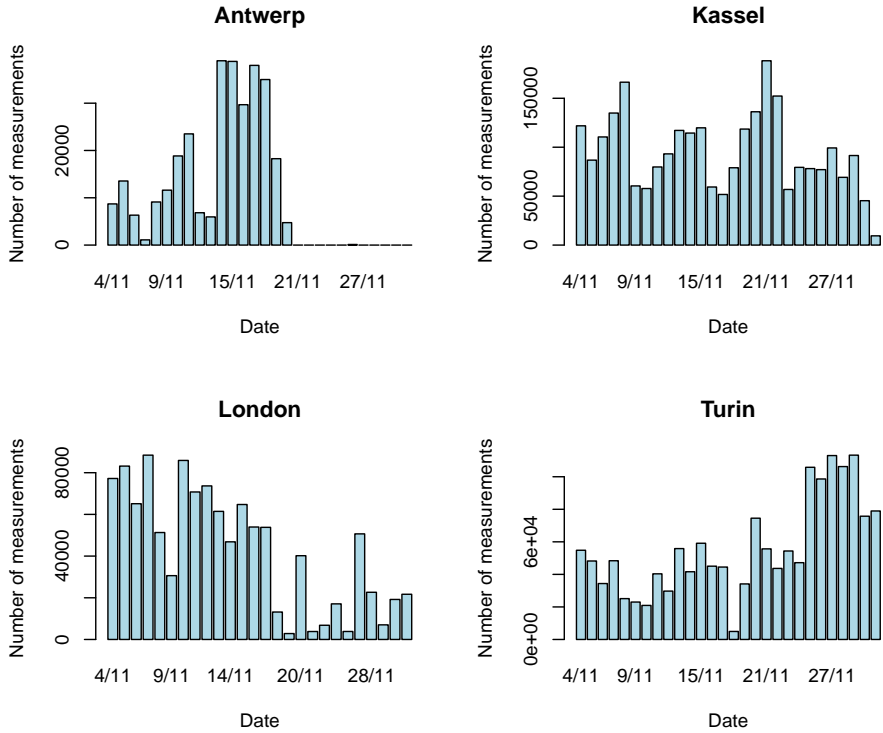

Figure 15: Number of data points per day in the four locations of the challenge.

the different locations, in Table 8, in an aggregated manner: for each location we show the total number of 10 m ×10 m squares covered (space coverage) and the average number of measurements (rough measure of time coverage). The table shows that coverage follows the same trend as the number of measurements (Table 7): the highest coverage is achieved by Kassel, which also won the challenge, and lowest by Antwerp. This applies both to the space dimension (surface covered at least once) and time (number of repeated measurements in an area). In total, volunteers covered over 24 km$^2$, and each 10 m ×10 m tile contained on average 24 measurements.

|  | Europe | Antwerp | Kassel | London | Turin |
| --- | --- | --- | --- | --- | --- |
| Surface covered in $m^2$ | 24,330,700 | 1,906,500 | 8,373,400 | 6,996,000 | 7,054,800 |
| Average number of measurements per $100m^2$ | 24.14 | 17.01 | 34.71 | 15.36 | 22.23 |

Table 8: Coverage obtained in Europe and at each of the four locations.

For an improved qualitative image of the type of coverage patterns in the different weeks of the

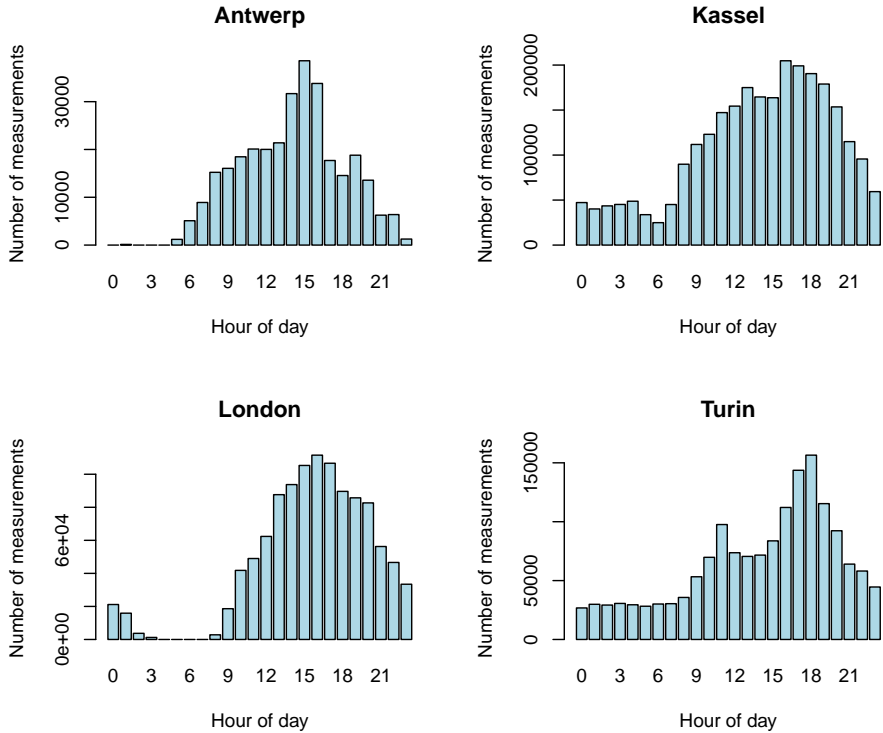

Figure 16: Number of data points per hour of the day at the four locations of the challenge.

challenge, we provide two examples from two different teams in Turin (Figures 17 and 18). These validate the observations made in the main text: during the first two weeks of measurements teams explore more, in their aim to cover the area of interest, i.e., the predefined mapping area, as well as possible. In the last phase, however, when no mapping area exists, they perform repeated measurements on their daily tracks, with reduced space exploration. This pattern is important for further analyses, since the space/time coverage appears to be much better when the area is restricted.

For a better view of the evolution of measured pollution levels between phases 2 and 3, Figure 19 shows the distribution of BC for the different locations, compared in the different phases. We use notched boxplots, which show minimum, maximum and quartile values for the data: the box represents the range of the data between the first and third quartile. The notches show confidence intervals (if these do not overlap, differences between the distributions are significant). In the plots presented here, the notches are so small that they are invisible. The plots also contain information about the size of the different datasets: the width of the boxes is proportional to the square root of the number of data points represented. In Kassel, volunteers were grouped into two groups in phase 3: the first group (g1 - three sensor boxes) had as a task to avoid highly polluted areas, while group g2 (6 sensor boxes) had no task other than using the sensor box where they wished. This in order to test whether any learning appears during measurements.

For Antwerp, volunteers collected much higher BC levels in phase 3. In London, although

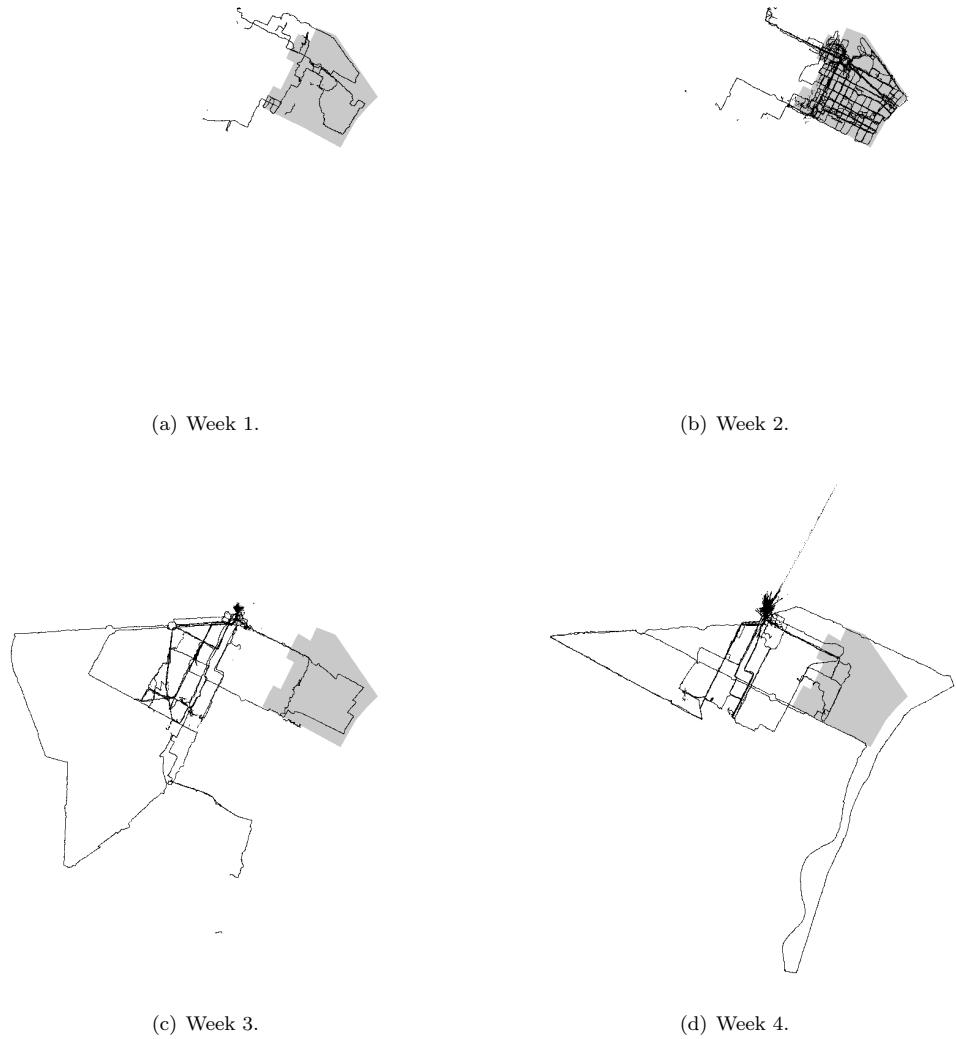

(a) Week 1.  (b) Week 2.

(c) Week 3.  (d) Week 4.

Figure 17: **Coverage for group "ggwp"** over the four measuring weeks of the APIC challenge. The grey area indicates the predefined mapping area.

means are not larger, the maximum levels achieved are larger in phase 3. However, for these two cities data in phase 3 is rather limited compared to the other locations and to phase 2 (as shown by the width of the boxplots in Figure 19 and in Figure 15). For Turin, an increase in the measured pollution levels is clear again. So, for all three locations, there is a good indication that

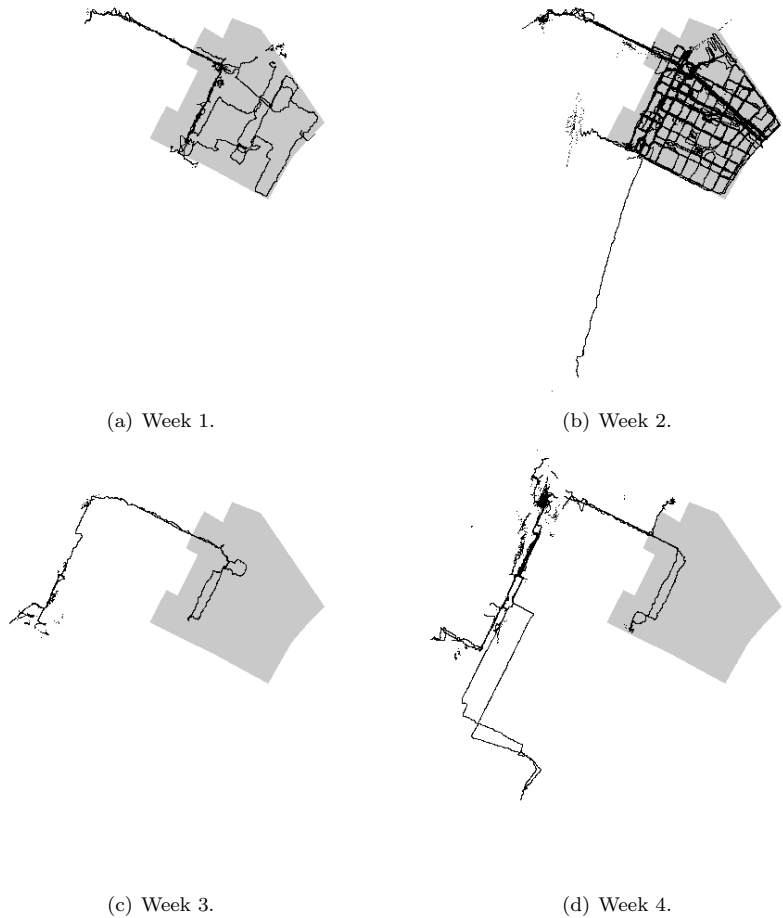

(a) Week 1.  (b) Week 2.

(c) Week 3.  (d) Week 4.

Figure 18: **Coverage for group "TUX"** over the four measuring weeks of the APIC challenge.

volunteers concentrated more on high pollution levels in the 3rd phase of the challenge: when they were allowed to explore, the aim was to identify highly polluted locations.

For Kassel, the group tasked with minimising their exposure (g1) displays on average larger BC levels compared to the other group. Maximum values appear, however, to be lower compared both to the previous phase and to g2. This may indicate that volunteers have only learned how to avoid extreme pollution levels, but still cannot discriminate when it comes to average behaviour.

Of course, pollution levels themselves may change from one day or period to another, making evaluation of user behaviour difficult. For instance, if a user appears to measure higher values in time, this could be either because of a shift in his personal interests, or because pollution itself increases. For the period of the challenge, PM10 data (particulate matter with an aerodynamic diameter under 10 $\mu$m) were available from official monitoring stations for all locations, while BC only for Antwerp [8, 9, 10, 11]. Although BC is mainly represented in the small diameter ranges

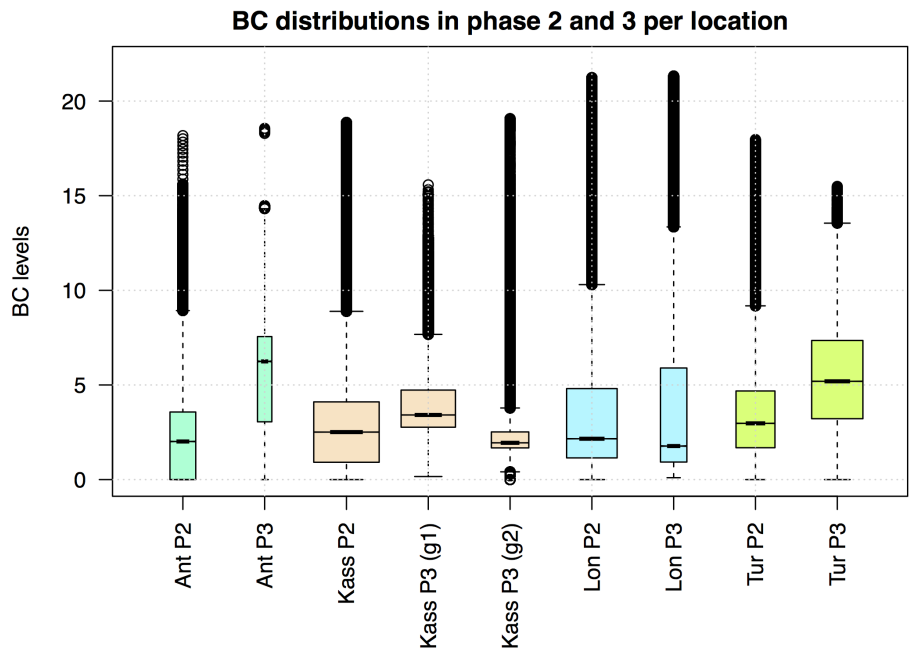

Figure 19: **Pollution levels per location compared in the two phases.** The distribution of BC levels (in $\mu g/m^3$) are shown for the two measuring phases of the challenge, phases 2 and 3, separate for each location. For phase 3 Kassel's AirAmbassadors are split into two groups with different objectives. Group 1 (g1) was supposed to avoid strongly poluted areas. Group 2 (g2) had no specific goal.

of PM, PM10 data were used for comparison with the measurements, due to higher availability, since they give a good indiction of the general level of pollution. In Figure 20, we compare the daily average PM10 with average BC values obtained by our volunteers. In Antwerp, we also show official average BC levels. As the figure shows, BC levels measured by our volunteers are within a good range compared to PM10 values. Differences are comparable to those observed between reference BC and PM10 in Antwerp and may indicate some particular interest of the volunteers. Also, no increase between phases 2 and 3 is visible in PM10 data. Table 9 shows average PM10 for all locations for phases 2 and 3 (for Antwerp we also show BC). This confirms that no significant increase in overall pollution levels appeared from phase 2 to phase 3.

Table 9: Average official pollution levels for the four locations in phase 2 and phase 3 of the challenge.

|         | Antwerp PM10 | Antwerp BC | Kassel PM10 | London PM10 | Turin PM10 |
|---------|--------------|------------|-------------|-------------|------------|
| Phase 2 | 29.25        | 4.33       | 26.2        | 26.73       | 50.42      |
| Phase 3 | 28.29        | 3.03       | 28.1        | 26.05       | 39.66      |

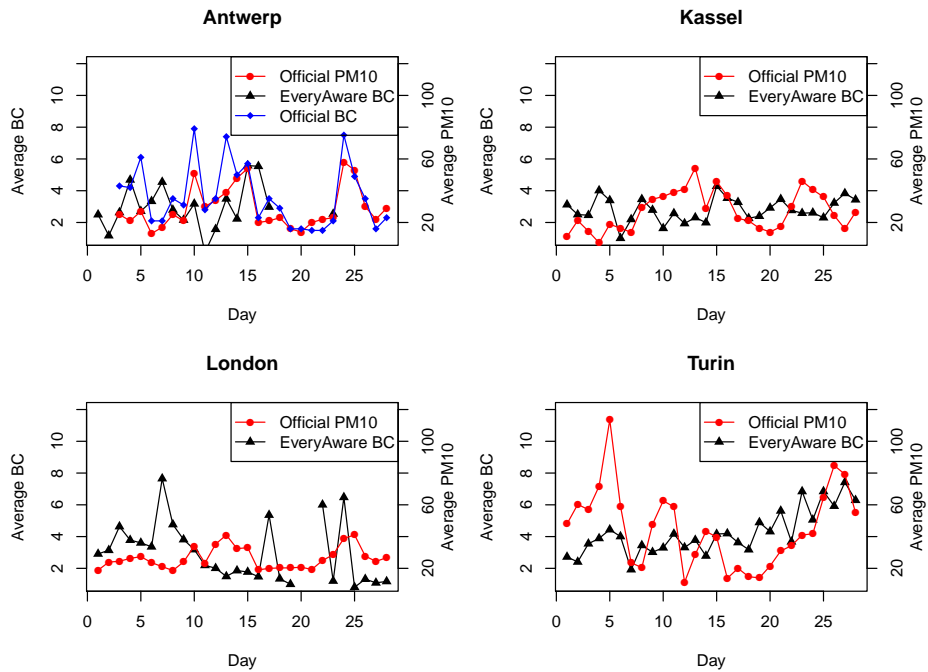

Figure 20: Comparison of measured average BC to PM10 and BC reference measurements in the areas of interest.

## 7.1 Opinion evolution model

This section revisit the APD graphs from the main text aggregated according to each phase reported in Figure 21.

28

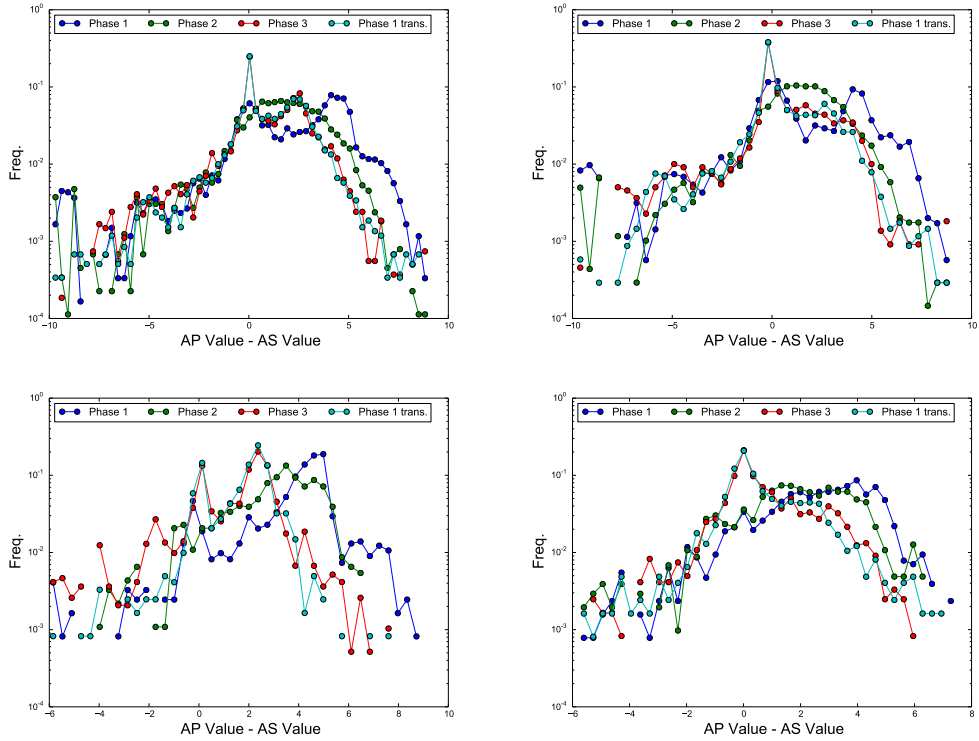

Figure 21: Clockwise, from the top left: the APD histogram for the overall, for Kassel, for Turin and for London in each phase of the challenge and with an estimation of phase 3 data obtained from phase 1 data through the transformation defined in Eq. (5).

29

If we look at phase 3 histograms two main features attract our attention: a narrow peak at 0 and a strongly asymmetric structure. The first feature was somehow expected since players are trusting the AS (AirSquare) values shown in the AS, and they are annotating accordingly. Fortunately, the peak at zero is not delta like, which would be expected if users were copying the AS value. Rather, players still have their opinion on the environment and keep it despite the on field measurements. This may happen because they are really trying to follow the basic ideas of the game, but also because copying it is not the best strategy, since the AS value is aggregated, i.e. it is the average of all sensor box measurements taken in the corresponding AS, while the real measurements used for revenue calculation were punctual values which could be substantially different. So the shape of the distribution around zero seems to be caused by users learning the most likely air quality value and trying to estimate fluctuations. But graphs in Figure 21 show something more. There is a clear asymmetry for phase 3 distributions, since the great part of APD values fall in the positive range. This could be a consequence of the fact that AS values were around 3 $\mu g/m^3$ so there was a 30% probability to underestimate that value and 70% to over estimate, but if we look at the phase 1 distributions, this asymmetry effect seems better explained by a sort of memory effect or inertia of players in changing their opinions. This hypothesis seems realistic if we look at the London graph. The main peak around 4 $\mu g/m^3$ is still present in phase 3, although it is shifted. In order to measure this effect we defined a transformation that takes into the account both features just discussed: the accumulation around 0 and the shift. Let us consider a given set of opinions $o_i$ about a certain number of topics provided by a certain number of subjects. At a given time those subjects are exposed to values $h_i$, which are perceived as hints of the true values. We are interested in what happens to the difference between opinions and hints before and after the exposition, to understand how this information will affect the opinion structure. To this aim, we define the set of differences $d_i$ between the opinions and the relative hints and analyse the distribution of those differences before and after the exposition. Obviously, the variation of the differences is only due to the variation of the opinions. As we said, we want to reproduce the phenomenon of the accumulation around the hints (i.e., $d_{aft} \sim 0$) and the shift of the general opinion, that we will try to describe as a sort of rescaling (i.e., $d_{aft} \sim d_{bef}/r$ where $r$ will be the rescaling constant). Which of the two phenomena will take place will be decided randomly: with a given probability $p_0$ the opinion will reset around 0, otherwise, with probability $1 - p_0$, the opinion will just be rescaled. Finally, around this two attractors we add a certain amount of noise. We decided for a Cauchy distribution $C(X)$ centered at 0 in one case and at $d_{bef}/r$ in the other, i.e.

$$C(x; \mu, \gamma) = \frac{1}{\pi \gamma \left(1 + \left(\frac{x-\mu}{\gamma}\right)^2\right)} \tag{4}$$

where $\mu$ is the average (and the center of this symmetric distribution) and $\gamma$ represents a scale factor. It is worth to note that the variance of this distribution is not defined, since the second momentum of the distribution does not converge. This choice seems reasonable because tails seem to be power law-like rather than gaussian-like, as the log plots in Figure 21 show. Let us define our transformation and its effect on the difference $d_{bef}$ between the opinion and the hint before the exposure. According to the rules we stated earlier, $d_{aft}$ will be distributed according to this density function:

$$T(d_{aft}; d_{bef}, p_0, r, \gamma_0, \gamma_r) = \begin{cases} C(d_{aft}; 0, \gamma_0) & \text{with prob. } p_0 \\ C(d_{aft}; d_{bef}/r, \gamma_r) & \text{with prob. } 1 - p_0 \end{cases} \tag{5}$$

The transformation we just defined introduces four parameters:

- $p_0$, which is the probability that the old opinion is reset around $d = 0$; thus, with probability $1 - p_0$, the opinion shows a certain inertia; this resistance to change causes a shift toward the hint instead of a complete reset;

- $r$, the rescale factor quantifying the shift of resilient opinions;

- $\gamma_0$ and $\gamma_r$, the $\gamma$ scale factors for the Cauchy distributions centered respectively at 0 and at $d_{bef}/r$ introduced to add a realistic noise.

We used our data to infer the parameters of our model for Kassel, London, Turin and for the complete set of data. If we apply the transformation to phase 1 data, we get an estimate of phase 3 distances between opinions and hints. Then, to evaluate how good is the estimate, we use a two sample Kolmogorov-Smirnov two sided test. This kind of test gives as result the probability $p_{val}$ that the hypothesis that the two samples are drawn from the same distribution cannot be rejected. Usually, a value below 5% means that the hypothesis has to be rejected otherwise the hypothesis is likely to be true. If the $p_{val}$ is around 10% the two samples come from two distribution which are, in any case, very close. Above 30% the samples can be considered with a good degree of confidence as coming from the same distribution. We explored the space of parameters with 10% steps and repeating the test 100 times to find the combinations with the highest $p_{val}$ for Kassel, London, Turin and for the overall. These optimal combinations are reported in Table 10 with the relative results for the Kolmogorov-Smirnov test.

Table 10: Parameter combinations with the highest $p_{val}$ resulting from the Kolmogorov-Smirnov test. Parameter space has been explored with 10% steps and each configuration has been tested 100 times. The average $p_{val}$ is reported. Some peaks in the tails for London compromised the test, causing as a result unsatisfying values for the parameters. We reduced the range in the most meaningful area, which is $(-1 : 4]$). We found the best parameters testing only this area, obtaining a remarkable result ($p_{val} = 27\%$). Then we made again the test reintroducing neglected data, obtaining a $p_{val} = 9\%$ which is still a satisfactory result.

| dataset | $p_0$ | $r$ | $\gamma_r$ | $\gamma_0$ | $< p_{val} >$ |
|---|---|---|---|---|---|
| Kassel | 0.336 | 1.62 | 0.381 | 0.0138 | 0.192 |
| London | 0.147 | 1.90 | 0.100 | 0.030 | 0.267 (0.087) |
| Turin | 0.583 | 1.56 | 0.304 | 0.300 | 0.417 |
| Overall | 0.204 | 1.767 | 0.28 | 0.015 | 0.262 |

From Table 10 it appears that the reset of the opinion around the hint does not happen so often. In London, for example, it is almost a secondary effect. In the *best* case, Turin, the reset seems to be there slightly more than in half of the cases. We also reported in Figure 21 an estimate of the APDs for phase 3 obtained by applying the transformation 5 with the *optimal* parameter combination to the data of phase 1. The similarity between the estimate and phase 3 real data is pretty clear.

It is very likely that Eq. (5) is not the real transformation of the opinion due to the subjects' exposure to hints. We made strong assumptions and we reduced our data set to focus on the interesting part. Also, we are analyzing and modeling the phenomenon on a very narrow time scale (weeks) without knowing almost anything about the others (for example, if we considered months the dynamics could be potentially extremely different). Despite these considerations, the results we showed point out with sufficient reliability that the main components are there. The model we referred to helped us to measure how our volunteers were influenced by the hints we gave them. We may now affirm with a certain degree of confidence that even when people do not trust completely the AS values, they still get influenced by them. Another way to see this is that, even if people do not reset their opinions, the space itself in which their opinions are

arranged is deformed by the exposure to hints. Obviously these considerations are justified if the subjects consider the source of the hints as *objective*. In other cases, for example, if volunteers are told that opinions come from other volunteers, completely different dynamics are expected to come into play.

# References

[1] EPA. Report to congress on black carbon - external peer review draft. Tech. Rep. EPA-450/D-11-001, EPA (2010).

[2] Mitchell, T. M. Machine learning. 1997. *Burr Ridge, IL: McGraw Hill* **45** (1997).

[3] Hagler, G., Yelverton, T., Vedantham, R., Hansen, A. & Turner, J. Post-processing method to reduce noise while preserving high time resolution in aethalometer real-time black carbon data. *Aerosol and Air Quality Research* **11**, 539 – 546 (2011).

[4] Google. http://www.google.com/permissions/geoguidelines/attr-guide.html (Date accessed: 03/02/2015).

[5] Becker, M., Mueller, J., Hotho, A. & Stumme, G. A generic platform for ubiquitous and subjective data. In *1st International Workshop on Pervasive Urban Crowdsensing Architecture and Applications, PUCAA 2013, Zurich, Switzerland – September 9, 2013. Proceedings*, 1175–1182 (ACM, New York, NY, USA, 2013).

[6] http://www.openstreetmap.org/copyright (Date accessed: 03/02/2015).

[7] Caminiti, S. *et al.* Xtribe: A web-based social computation platform. In *Cloud and Green Computing (CGC), 2013 Third International Conference on*, 397–403 (2013).

[8] The Flemish Environment Agency. Antwerp air quality data, http://www.vmm.be/ (date of access: 01/02/2014).

[9] Environmental Research Group, King's College London. London air quality data, London Air, http://www.londonair.org.uk/london/asp/datadownload.asp, (Date of access: 25/01/2014).

[10] Comune di Torino. Turin air quality data, Agenzia Regionale per la Protezione Ambientale del Piemonte, http://www.comune.torino.it/ambiente/inquinamento/new-report.php, (Date of access: 25/01/2014).

[11] Kassel air quality data, Hessisches Landesamt fur Umwelt und Geologie, http://www.hlug.de/?id=7122&view=messwerte&detail=download&station=1415, (Date of access: 25/01/2014).


\end{document}